\documentclass[%
 reprint,
  superscriptaddress,
 amsmath,amssymb,amsfonts,
 aps,
 prl,
 floatfix,
 longbibliography
]{revtex4-2}

\usepackage{dcolumn}
\usepackage{bm}

\usepackage[pdftex]{graphicx} \graphicspath{{}}
\usepackage{float} \usepackage{color}

\usepackage[pdftex,colorlinks=true]{hyperref}
\hypersetup{
  colorlinks=true,
  linkcolor=blue,
  urlcolor=cyan,
}

\usepackage{mathtools}

\DeclarePairedDelimiter\ket{\lvert}{\rangle}
\DeclarePairedDelimiterX\braket[2]{\langle}{\rangle}{#1 \delimsize\vert #2}
\DeclarePairedDelimiterX\expval[3]{\langle}{\rangle}{#1 \delimsize\vert #2  \delimsize\vert #3}

\DeclarePairedDelimiterX\sexpval[1]{\langle}{\rangle}{#1}

\usepackage{siunitx}

\newcommand{\vect}[1]{\mathbf{#1}}

\def\sectionn#1{\noindent\underline{\it #1:}}

\begin{document}

\title{Dynamical generation of spin squeezing in ultra-cold  dipolar molecules}

\author{Thomas Bilitewski}
\affiliation{JILA, National Institute of Standards and Technology and Department of Physics, University of Colorado, Boulder, CO, 80309, USA}
\affiliation{Center for Theory of Quantum Matter, University of Colorado, Boulder, CO, 80309, USA}
\author{Luigi De Marco}
\affiliation{JILA, National Institute of Standards and Technology and Department of Physics, University of Colorado, Boulder, CO, 80309, USA}
\author{Jun-Ru Li}
\affiliation{JILA, National Institute of Standards and Technology and Department of Physics, University of Colorado, Boulder, CO, 80309, USA}
\author{Kyle Matsuda}
\affiliation{JILA, National Institute of Standards and Technology and Department of Physics, University of Colorado, Boulder, CO, 80309, USA}
\author{William G. Tobias}
\affiliation{JILA, National Institute of Standards and Technology and Department of Physics, University of Colorado, Boulder, CO, 80309, USA}
\author{Giacomo Valtolina}
\affiliation{JILA, National Institute of Standards and Technology and Department of Physics, University of Colorado, Boulder, CO, 80309, USA}
\author{Jun Ye}
\affiliation{JILA, National Institute of Standards and Technology and Department of Physics, University of Colorado, Boulder, CO, 80309, USA}
\author{Ana Maria Rey}
\affiliation{JILA, National Institute of Standards and Technology and Department of Physics, University of Colorado, Boulder, CO, 80309, USA}
\affiliation{Center for Theory of Quantum Matter, University of Colorado, Boulder, CO, 80309, USA}

\date{\today}

\begin{abstract}
We study a bulk fermionic dipolar molecular gas in the quantum degenerate regime confined in a two-dimensional geometry. 
Using two rotational states of the molecules we encode a spin 1/2 degree of freedom. 
To describe the many-body spin dynamics of the molecules we derive a long-range interacting XXZ model valid in the regime where motional degrees of freedom are frozen. 
Due to the spatially extended nature of the harmonic oscillator modes, the interactions in the spin model are very long-ranged and the system behaves close to the collective limit, resulting in robust dynamics and generation of entanglement in the form of spin squeezing even at finite temperature and in presence of dephasing and chemical reactions. We discuss how the internal state structure can be exploited to realise time-reversal and enhanced metrological sensing protocols.
\end{abstract}

\maketitle
\sectionn{Introduction}
Systems of dipolar molecules \cite{Review_dipolar,Review_molecules} have been shown to be versatile simulators of long-range quantum spin models \cite{Micheli2006,PhysRevLett.96.190401,AlexeyPRA,AlexeyPRL,KadenPRL,KadenPRL2},
with prospects ranging from the study of fundamental physics \cite{DeMille990} to applications in quantum devices \cite{Andre2006} and quantum metrology \cite{RevMod_Metrology_2018}. 

While the complex internal structure of molecules makes these systems particularly attractive, it also results in inelastic lossy collisions \cite{Ni2010,PhysRevX.8.041044,Ospelkaus853,Gregory2019}.
A lot of progress has been made in 3D optical lattices \cite{PhysRevLett.98.060404,deMiranda2011,PhysRevLett.108.080405,Moses659}, where these losses are suppressed, from the demonstration of   long-range spin exchange \cite{Spin_exchange_nature_2013} and control over the interactions \cite{KadenPRL2} to the study of Zeno suppression \cite{Zhu2014}. 
However, these studies have been limited to non-degenerate gases which suffer  additional heating mechanisms when loaded  into an optical lattice.
The recent realization of a  quantum degenerate gas of fermionic  molecules in  a bulk system \cite{DeMarco2019,PhysRevLett.124.033401}, where chemical reactions  inherent to dipolar molecules \cite{PhysRevLett.124.163402,Hu1111,Kirste1060,PhysRevLett.123.123402} can be suppressed by confining the gas to two dimensions \cite{deMiranda2011,PhysRevLett.105.073202,PhysRevA.83.012705,PhysRevA.88.063405,2007.12277v1} opens untapped  opportunities. These include the exploration of  many-body physics with  tunable elastic long-range dipolar interactions in regimes not accessible before.

\begin{figure}
\includegraphics[width=\columnwidth]{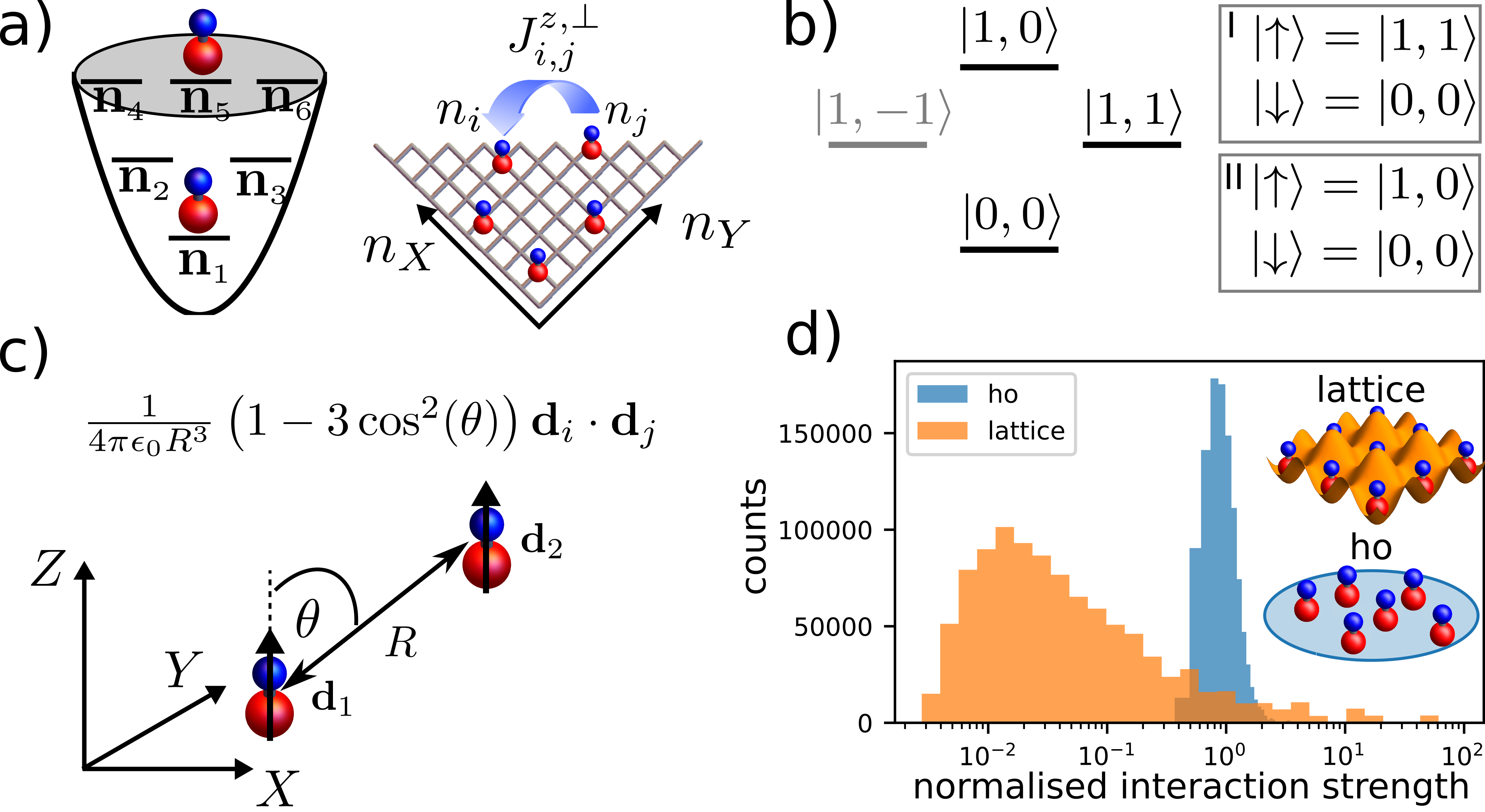}
\caption{%
    a) Dipolar molecules confined in a 2D harmonic trap in oscillator modes $n=(n_X,n_Y)$ are mapped to a XXZ spin model in mode space. %
    b) Internal rotational levels $\ket{\mathcal{N},\mathcal{N}_Z}$ and the two different spin-bases used in this work.
    c) Dipolar molecules interact via long-range $1/R^3$ dipole-dipole interactions. %
    d) Histogram of all pairwise interactions normalised to the mean interaction strength for $N=1000$ molecules in a single plane of a 3D optical lattice ('lattice') or in harmonic oscillator ('ho') modes of a 2D harmonic trap.\label{fig:illustration}}
\end{figure}
Here, we study the   dynamics of a dipolar molecular gas  prepared in the quantum  degenerate  regime  and  confined in a  two dimensional harmonic potential with two relevant rotational levels that form  an effective  spin 1/2 degree of freedom. 
We derive a long-range interacting XXZ spin model describing the many-body dynamics of this system in the regime where molecules remain frozen in the harmonic oscillator modes.
We point out 3 major advantages of these systems. Firstly, the 
quasi-2D  confinement enhances elastic interactions and   protects the molecules against undesirable  chemical reactions \cite{2007.12277v1}. Secondly, the spatially extended nature of the motional states results in a very long-range spin model which features   spin dynamics robust to thermal noise, dephasing, and s-wave losses. In fact, the model is very close to the one-axis twisting model \cite{Kitagawa1993}, known to produce spin-squeezed states useful for quantum metrology \cite{Entanglement_OAT_2009,RevMod_Metrology_2018} as demonstrated in a variety of different platforms \cite{RevMod_Metrology_2018,Sorensen_OAT_BEC,Vladan_2010a,Vladan_2010b,Ions_Bollinger_2016,Vladan_Clock_2020}. %
Indeed, we predict up to $19$ dB of spin-squeezing with 1000 molecules.
Finally, time-reversal can be realised by tuning an applied electric field, or by state transfer between rotational molecular levels, allowing for  the implementation of robust metrological  protocols  for precise electromagnetic field sensing, that fully take advantage of entanglement without the need of single photon detection capabilities \cite{Schleier_Smith_Heisenberg,Schulte_2020}.

\sectionn{Model} We now turn to deriving the spin model for dipolar fermionic molecules in quasi-2D occupying harmonic oscillator states and interacting via long-range dipolar interactions as illustrated in Fig.~\ref{fig:illustration}a)-c).
 
The effective spin 1/2 degree of freedom is encoded in the internal rotational levels of the molecules. We assume  coupling to  nuclear levels is suppressed, e.g. by a strong magnetic field \cite{AlexeyPRA}.
In this case, the level structure is described by the molecular rotor Hamiltonian in the  presence of an electric field, $\hat{H}_{rot} =  B \vect{\hat{N}}^2 - \hat{d}_0 E$ \cite{AlexeyPRA}, where $B$ is the rotational constant, $\vect{\hat N}$ the angular momentum operator of the molecule, $E$ the strength of the electric field oriented along the $Z$-direction, and $\hat{d}_0=\vect{\hat{d}}\cdot \vect{e}_Z$ the projection of the dipole operator along the field direction.
 The eigenstates $\ket{\mathcal{N},\mathcal{N}_Z}$ labelled by two rotational quantum numbers satisfy at vanishing field  $\vect{\hat{N}}^2\ket{\mathcal{N},\mathcal{N}_Z}=\mathcal{N}(\mathcal{N}+1) \ket{\mathcal{N},\mathcal{N}_Z}$ and $\hat{N}_Z \ket{\mathcal{N},\mathcal{N}_Z}= \mathcal{N}_Z \ket{\mathcal{N},\mathcal{N}_Z}$ where $\hat{N}_Z=\vect{\hat{N}}\cdot \vect{e}_Z$.
 In this work we will work with two distinct spin $1/2$ bases either $\ket{\downarrow}= \ket{0,0}$ and $\ket{\uparrow}=\ket{1,1}$ (basis I) or $\ket{\downarrow}= \ket{0,0}$ and $\ket{\uparrow}=\ket{1,0}$ (basis II) as shown Fig.~\ref{fig:illustration}b). %
 Note that quadrupolar interactions prevent coupling of these states to other  rotational levels allowing us to restrict the dynamics to only 2 levels. 
 
 Projected into this internal state basis the single particle Hamiltonian reduces to $\hat{H}_0 =\sum_i \mathcal{E}_{\alpha,i} {\hat c}^{\dagger}_{i,\alpha} \hat{c}_{i,\alpha}$, where $\hat{c}^{\dagger}_{i,\alpha}$ creates a fermionic molecule in internal state $\alpha=\uparrow,\downarrow$ and harmonic oscillator mode $i=(n^i_X,n^i_Y,n^i_Z)$ with energy $\mathcal{E}_{\alpha,i} = \mathcal{E}^{rot}_{\alpha} +\hbar (\omega_{\alpha,X} n^i_X + \omega_{\alpha,Y}n^i_Y +\omega_{\alpha,Z} n^i_Z)$. We assume isotropic confinement within the plane, $\omega_\alpha=\omega_{\alpha,X}=\omega_{\alpha,Y}$, and the confinement along $Z$ to be the largest energy scale, larger than the Fermi Energy, $\epsilon_F$, and the thermal energy, $k_{\rm B}T$, 
 such that molecules only occupy the corresponding ground state, $n^i_Z=0$.

We express the dipolar interactions in this basis as
\begin{equation}
     1/2 \sum_{ijkl} V_{ij}^{kl}
   \sum_{\alpha\beta} \mu_{\alpha} \mu_{\beta} \hat{f}^{lk\dagger}_{\beta \alpha}\hat{f}^{ij}_{\beta \alpha} +  \mu_{\downarrow\uparrow}\mu_{\uparrow\downarrow} ( \hat{f}^{lk\dagger}_{\uparrow\downarrow} \hat{f}^{ij}_{\downarrow\uparrow}  +h.c.)
   \label{eq:int}
\end{equation}
where we ignored the dependence of the spatial modes on the internal molecular state. Here  $V_{ij}^{kl}= \expval{ij}{{\hat V}_{dd}}{kl}$ with $\langle \vect{R}|\hat{ V}_{dd}|\vect{R}\rangle  = \frac{1}{4\pi \epsilon_0 R^3} (1-3 \cos^2(\theta))$, and $\theta$  the angle between the  vector connecting the pair of interacting molecules  $\vect{{R}}$ and  $\vect{e}_Z$ (see Fig.1). We used the abbreviation $\hat{f}^{ik}_{\alpha \beta}=\hat{c}_{i\alpha} \hat{c}_{k\beta}$, and defined the dipole moments, $\mu_{\alpha} = \langle\alpha|\hat{d}_0|\alpha
\rangle$, and $\mu_{\uparrow \downarrow}=\mu_{\downarrow \uparrow}=\langle\uparrow\rvert\hat{d}_0\lvert\downarrow\rangle$ for basis II, $\mu_{\downarrow \uparrow}=\langle \downarrow \rvert \hat{d}_- \lvert \uparrow\rangle/\sqrt{2}$, $\mu_{\uparrow\downarrow}=\langle \uparrow \rvert \hat{d}_+ \lvert \downarrow\rangle/\sqrt{2}=-\mu_{\downarrow\uparrow}$ for basis I, with the spherical components $\hat{d}_{0,\pm}$ of $\hat{\vect{d}}$. 

In the collision-less regime in which the internal spin dynamics is faster than collisional processes relaxing the motional degrees of freedom \cite{supplemental,Deutsch_2010} and assuming at most one molecule per mode (achievable by initializing a spin polarized gas), interaction induced mode changing processes can be neglected \cite{Smaleeaax2019,martin2013} and only  couplings between states at the same single-particle energy, e.g. $i=k$, $j=l$ or $i=l$, $j=k$, need  to be kept in Eq.~\ref{eq:int} to leading order. In this limit, the Hamiltonian  can be reduced  to a  long-range interacting XXZ spin model \cite{supplemental}
\begin{align}
  \mathcal{H} &= 1/2 \sum_{ij} J^z_{ij} \hat{s}^z_{i} \hat{s}^z_{j}  + J^{\perp}_{ij}  \left( \hat{s}^x_{i} \hat{s}^{x}_{j}+ \hat{s}^{y}_{i}\hat{s}^{y}_{j} \right)  + \sum_{i} \hat{s}^z_{i}\, h^z_{i}
  \label{eq:spin_model}
\end{align}
where $\hat{s}_{i}^{\nu}=1/2\sum_{\alpha,\beta} \hat{c}^{\dagger}_{i\alpha} \sigma^{\nu}_{\alpha\beta} {\hat c}_{i\beta}$ are pseudo-spin 1/2 operators defined via the Pauli matrices $\sigma^{x,y,z}$. The spin couplings are given by $J^z_{ij} = \eta V_{ij}^{ji}   -  (\nu-\zeta)V_{ij}^{ij}$, $J^{\perp}_{ij}=  (\eta-\nu) V_{ij}^{ij}+\zeta V_{ij}^{ji}$ and $h^z_{i}   = \eta \sum_{k} 
  \left( V_{ik}^{ki}-V_{ik}^{ik}\right)/2 + \Delta \mathcal{E}_i$ 
with $\eta=(\mu_{\downarrow}-\mu_{\uparrow})^2$, $\nu =(\mu_{\downarrow}+ \mu_{\uparrow})^2$, $\zeta=2\mu_{\downarrow\uparrow}\mu_{\uparrow\downarrow}$, and  $\Delta \mathcal{E}_{i} = \mathcal{E}^{rot}_{\uparrow} - \mathcal{E}^{rot}_{\downarrow} + \hbar (\omega_{\uparrow} - \omega_{\downarrow})(n^i_X+n^i_Y)$.

\sectionn{Interactions in mode space}
We next discuss the form of the interactions in the spin model for spatially delocalised molecules in a harmonic trap compared to spatially localised molecules in deep real space lattices. 
We first note that in contrast to localised Wannier orbitals for which the terms $V_{ij}^{ij}$ are exponentially suppressed \cite{AlexeyPRA}, they are non-negligible for harmonic oscillator eigenmodes. In particular, the finite $V_{ij}^{ij}$ terms lead to a non-vanishing $J_z$ term even at zero-applied electric field, where $\mu_{\downarrow}=\mu_{\uparrow}=0$,  which is absent in the lattice system. 

To study the interaction between modes $i,j$
in more detail we consider $V=V_{ij}^{ij}-V_{ij}^{ji}$.  Explicit numerical evaluation shows this to decay only very slowly, see Fig.~S3 \cite{supplemental}, and a semiclassical calculation  \cite{supplemental} predicts significantly weaker scaling than for real space interactions which decay as $R^{-3}$ with the distance $R$. 
To visualise all resulting interactions in the spin-model and show the advantage of working with spatially delocalised molecules, we consider two physically distinct scenarios: a unit filled 2D array of N dipoles, localised at lattice sites $i=(i_X,i_Y)$, or a 2D harmonic trap, where dipoles occupy modes $i=(n^i_X,n^i_Y)$ up to the Fermi-level.
We show the resulting distribution of all pair-wise interactions in Fig.~\ref{fig:illustration}d).
We observe a wide distribution of couplings spanning many orders of magnitude for the real space lattice, compared to a sharply peaked distribution for the harmonic trap.
This small variance of couplings is key to the collective nature of the spin model facilitating robust spin dynamics. 

\sectionn{Collective limit}
Given this weak mode-dependence much of the physics of the spin model can be understood by considering the fully collective limit. Defining collective spin operators $\hat{S}_{\alpha}=\sum_i
\hat{s}^{\alpha}_i$ and the averaged couplings $\bar{J}_{\alpha}
=\frac{1}{N^2}\sum_{i,j} J^{\alpha}_{ij}$  and $ \bar{h}_z= \frac{1}{N}\sum_{i}
h^z_i$ we obtain a one-axis twisting Hamiltonian \cite{Kitagawa1993}
\begin{equation}
  \mathcal{H}_c =  \bar{J}_{\perp} \hat{S}^2 + \chi \hat{S}_z^2 +\bar{h}_z \hat{S}_z
\end{equation}
with $\chi \equiv\bar{J}_z - \bar{J}_{\perp}=
\mu(E) (\bar{V}_{ij}^{ij} - \bar{V}_{ij}^{ji})$ with $\mu(E)\equiv \left( -(\mu_{\downarrow}-\mu_{\uparrow})^2+2 \mu_{\downarrow\uparrow}\mu_{\uparrow\downarrow} \right)$. We note that through the dipole moments the interactions depend both on the electric field and the chosen set of coupled rotational states. 
In particular, by choosing either basis I or II we obtain a factor of $-2$ in the effective interactions \cite{Buechler2008,AlexeyPRA}, allowing us to reverse the dynamics. 

\sectionn{Parameters and methods} 
For specificity and to make predictions of value to near-future experiment, we specialise our calculations to dipolar KRb molecules \cite{DeMarco2019,2007.12277v1} and parameters accessible to current experiments: $\omega_Z=20$ kHz, $\omega=50$ Hz and distinct trapping frequencies of the internal states due to their AC polarisability set by $\Delta_\omega\equiv 2(\omega_{\uparrow}-\omega_{\downarrow})/(\omega_{\uparrow}+\omega_{\downarrow}) \approx 0.05 - 0.2$, \cite{Neyenhuis2012,2007.12277v1}. We present results for $N=100$ up to $1000$ molecules at temperatures ranging from $T/T_F =0$ up to $T/T_F=1.0$. The spin dynamics is obtained by solving the full spin model, Eq.~\ref{eq:spin_model}, via the  discrete truncated Wigner approximation
\cite{Schachenmayer2015,Zhu_2019} averaging over $10^4$ initial states and sampling the occupied modes from the  Fermi-Dirac distribution. We also include   $s$-wave losses from chemical reactions that  take place as the gas decoheres  \cite{supplemental}.

\sectionn{Robustness of dynamics to dephasing}
\begin{figure}
   \includegraphics[width=.99\columnwidth]{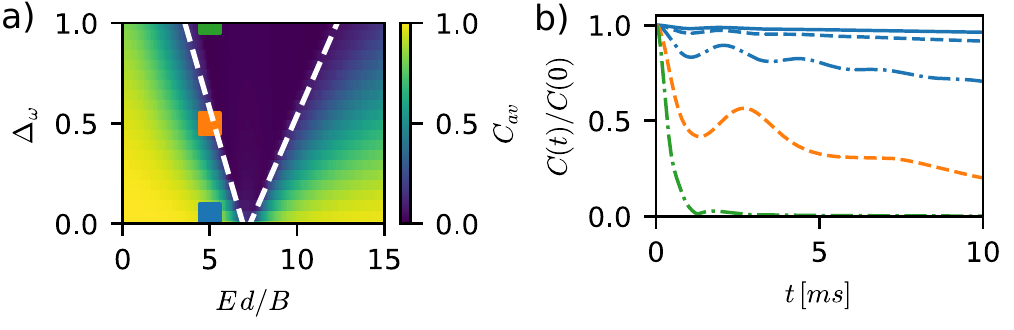}
  \caption{Dynamical phase transition. %
    a) Time-average contrast, $C_{av}=1/T_{av} \int_0^{T_{av}} dt \, C(t)/C(0)$ ($T_{av}=15$ ms), versus single-particle dephasing, $\Delta_\omega$, and electric field $E$ in units of $B/d$. White dashed lines indicate the transition. %
    b) Contrast $C(t)$ at fixed $E= 5 B/d$, $\Delta_{\omega}=0,0.05,0.02,0.5,1.0$ (top to bottom). %
    Mean field dynamics starting from a coherent state of $N=1000$ molecules along $x$ at $T/T_F=0$.
\label{fig:Fig2}}
  \end{figure}
We first  discuss the robustness of the dynamics to dephasing.
In the $\bar{J}_{\perp}=0$ limit,  the time-scale of dephasing is set by the standard deviation of the inhomogeneous $z$-fields proportional to $\Delta_\omega$. In this limit losses from chemical reactions  due to  $s$-wave collisions between molecules in the $\alpha=0$ and $\alpha=1(\tilde{1})$ states  play an important role too  \cite{supplemental}. 
In contrast, when $\bar{J}_{\perp}$ dominates the dynamics, the dephasing and $s$-wave losses are  strongly suppressed by  the opening of a many-body gap \cite{Rey2008} proportional to $ N \bar{J}_{\perp}$. The gap  facilitates spin locking  along the collective spin direction, a mechanism referred to as  spin-self rephasing \cite{Deutsch_2010}. The competition between dephasing and collective interactions has been shown to result in a dynamical phase transition (DPT) with two distinct dynamical behaviors as the system crosses a critical value of interaction strength $\bar{J}_{\perp}^c$ \cite{Smaleeaax2019}. 
The DPT is observed  by an abrupt change in the  the contrast $C(t)=\sqrt{S_x^2+S_y^2}$ at $\bar{J}_{\perp}^c$.
 Fig.~\ref{fig:Fig2}a) shows the long-time average of the contrast as a function of the dephasing term
$\Delta_\omega$ and applied electric field $E$ for an  initial coherent spin-state prepared in the $xy$ plane for an ideal system at zero temperature.  Note that  the energy gap  $\bar{J}_{\perp}$ depends on $E$, and  vanishes around $E\approx 7 B/d$ ($B/d=3.9 $ kV/cm  for KRb) \cite{supplemental}. Consequently, we observe robust interaction protected spin dynamics in the region of $|\bar{J}_{\perp}(E)|>\bar{J}_{\perp}^c(E,\Delta_{\omega})$ and an abrupt change to fast dephasing and subsequent chemical reaction losses  for $|\bar{J}_{\perp}(E)|<\bar{J}_{\perp}^c(E,\Delta_\omega)$ separated by a critical region indicated by the dashed line in Fig.~\ref{fig:Fig2}a). We illustrate the qualitatively different dynamics in Fig.~\ref{fig:Fig2}b) via time-slices at fixed electric field and dephasing $\Delta_{\omega}$ below, at, and above this transition. 

Besides single-particle dephasing and losses, the so far neglected mode changing collisions also disrupt the collective spin. To account for both single particle and interaction induced dephasing, we develop a kinetic model \cite{supplemental}. We find a decay time of the collective spin due to collisions of $\tau \approx 11$ ms at $T/T_F =1.0$, see Fig.~S4 in \cite{supplemental}, which is rapidly increasing at lower temperatures (Fig.~S5 in \cite{supplemental}), and thus, largely negligible for the time-scales of interest in the quantum degenerate regime.

\sectionn{Spin squeezing}
   \begin{figure}
   \includegraphics[width=.99\columnwidth]{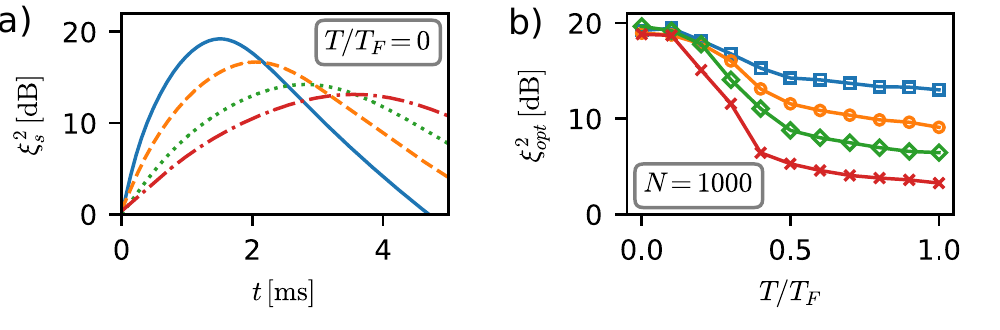}
  \caption{Dynamics and robustness of spin squeezing. a) Ramsey squeezing parameter $\xi_s^2$ versus time $t$ for $N=1000,400,200,100$ (solid, dashed, dotted, dashed-dotted) molecules. %
    b) Optimal Ramsey squeezing parameter $\xi_{opt}^2$
    versus temperature for different dephasing strengths $\Delta_\omega=0.0,0.05,0.1,0.2$ (squares, circles, diamonds, crosses). Both at zero electric field $E=0$.
  \label{fig:Fig3}}
  \end{figure}
Next we consider the generation of entangled many-body states during the time-evolution and their robustness to thermal fluctuations and dephasing.
In particular, we study the generated spin squeezing as characterised by the Ramsey squeezing parameter \cite{Wineland1992,Wineland1994}
\begin{equation}
  \xi_s^2= N \frac{\min_{\phi} \langle \mathrm{Var}[\hat{S}^{\perp}_{\phi}]\rangle }{\lvert\langle\hat{\vect{S}}\rangle\rvert^2}
  \end{equation}
which measures the minimal variance of spin noise distribution taken over all axes
parametrised by the angle $\phi$ perpendicular to the mean collective spin $\langle\hat{\vect{S}}\rangle$.
We focus on states initially prepared fully polarised along $+\vect{x}$
on the Bloch sphere. The squeezing dynamics at zero temperature without dephasing is shown in Fig.~\ref{fig:Fig3}a) and the optimal spin-squeezing in Fig.~\ref{fig:Fig3}b) for
experimentally realistic parameters of particle number, temperatures and
dephasing, with a maximal squeezing of
$\xi_s^2 \approx 19$ dB for $N=1000$ molecules at low temperatures.
This shows the robustness of the observed squeezing for molecules  in the quantum degenerate regime $T/T_F \lesssim 0.5$  for a broad range of single particle imhomogeneities.

\sectionn{Time-Reversal and robust sensing}
\begin{figure}
  \begin{minipage}[!t]{0.99\columnwidth}
    \includegraphics[width=.99\columnwidth]{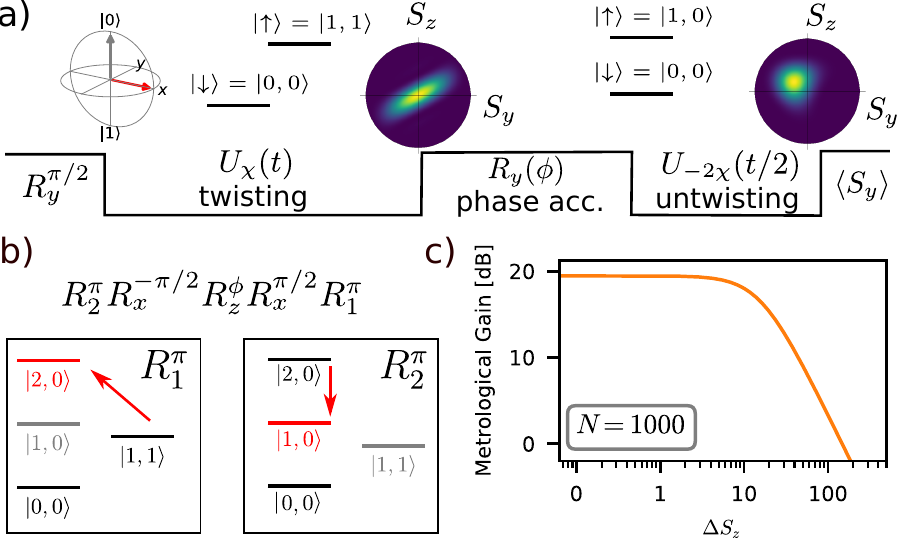}
  \end{minipage}
  \caption{Robust measurement via time-reversal. a) Illustration of the protocol: State preparation in a coherent state along x, evolution for time $t$, $U_{\chi}(t) =
    e^{i t \chi \hat{S}_z^2}$, resulting in a spin-squeezed state in the $\ket{0,0}$ and $\ket{1,1}$ basis, signal rotation $R_y(\phi)=e^{i \phi \hat{S}_y}$, state transfer from $\ket{1,1}$ to $\ket{1,0}$ realising $\chi \rightarrow -2 \chi$, evolution for $t/2$ 'untwisting' the
    state, followed by measurement of $\langle S_y \rangle$. Colour plots are Wigner functions for $N=10$ spins for the ideal protocol.
    b) Realisation of $R_y(\phi)$ via transfer to a non-interacting state $\ket{2,0}$, and accumulation of the phase due to free evolution via a compound sequence $R_y^{\phi}=\mathcal{R}_x^{-\pi/2} \mathcal{R}_z^{\phi}\mathcal{R}_x^{\pi/2}$
    c) Metrological gain $1/(N \Delta \phi^2) $ comparing the angular sensitivity
    $\Delta \phi$ to the standard quantum limit $1/\sqrt{N}$ versus measurement
    error $\Delta S_z$ computed from the full spin dynamics.
    \label{fig:time_reversal}}
\end{figure}
While in principle squeezed states are ideal for enhanced sensing, in practice, taking  full advantage of their enhanced sensitivity is  challenging due to  measurement noise limitations.  However, it has
been pointed out that by reverting the time-evolution and ``untwisting'' the state, it is possible to realize robust  Heisenberg limited phase sensitivity  without the need of single-particle-resolved state
detection \cite{Schleier_Smith_Heisenberg,Schulte_2020}. Below we discuss how to  implement the desired  ``untwisting''  protocol and robustly use polar molecules for precise  sensing of  electromagnetic fields.  

The basic protocol  consists of the following steps, illustrated in Fig.~\ref{fig:time_reversal}. After preparation of a coherent state along $x$, evolvution for a time $t$ by the dipolar Hamiltonian, $U_1 = e^{-i t \chi \hat{S}_z^2}$, generates a spin squeezed state.
This state is highly sensitive to rotations along the squeezed direction (which has a large projection along the $y$ axis). To perform precise measurements of a phase $\phi$ accumulated  under free evolution due to the energy difference of the internal states, which depends on external electromagnetic fields (see Fig.~S1 \cite{supplemental}),
 one just needs to align the state along the sensitive quadrature via $\mathcal{R}_x^{-\pi/2}\mathcal{R}_z^{\phi}\mathcal{R}_x^{\pi/2}$. To exclude undesirable dipolar interactions and many-body dephasing during phase accumulation one can first transfer the state $\ket{1,1}$ via a microwave $\pi$ pulse to $\ket{2,0}$  (or alternatively $\ket{0,0}$ to $\ket{1,0}$). This compound sequence to realise $R_y^{\phi}$ is illustrated in Fig~\ref{fig:time_reversal}b).  
The  ``untwisting'' protocol is performed by reversing  the dynamics $U_2 =e^{ i t \chi \hat{S}_z^2}$, followed  by a measurement  of $\langle \hat{S}_y\rangle $. Its final value is non zero due to the z-dependent spin precession induced by $\phi$.  
Time reversal at zero electric field can be  effectively accomplished in our system by coherently transfering all molecules in the $\ket{2,0}$ state to $ \ket{1,0}$  via  another  microwave $\pi$ pulse. Since 
 $\mu_{\uparrow\downarrow} \mu_{\downarrow\uparrow}  \rightarrow -2 \times \mu_{\uparrow\downarrow} \mu_{\downarrow\uparrow} $ letting the system evolve for $t/2$ reverses the dynamics.
 We additionally require $\pi$ pulses, $R_{y}^{\pi}$, at the middle of the twisting and untwisting steps to cancel inhomogeneous $z$-fields.

The advantage of untwisting protocols is the amplification of the spin rotation signal while keeping the quantum noise at the coherent state  level, $\Delta S_y \sim \sqrt{N}/4$. Therefore the sensitivity realised for a perfect noise-free measurement $\Delta \phi_0 = {\frac{\Delta S_y(\phi)}{\partial_{\phi}\left<S_y(\phi)\right>}}_{\phi=0} $ is only reduced by a factor $\sqrt{1+(\Delta  S_z/\Delta S_y)^2}$ in presence of measurement noise $\Delta S_z$ \cite{Schleier_Smith_Heisenberg}. In Fig.~\ref{fig:time_reversal}c) we show the metrological gain enabled by  the realisation of the protocol in our molecule system using  the full spin model, Eq.~\ref{eq:spin_model}. There we show the enhancement of sensitivity over the standard quantum limit versus final measurement noise in $S_z$, and observe the same gain as expected from an ideal implementation with a perfect unitary one-axis twisting dynamics.

\sectionn{Summary\label{sec:conclusions}} We have studied the spin dynamics of dipolar 
Fermi gases confined in a quasi-two-dimensional geometry, in regimes where losses can be effectively suppressed. By using delocalised eigenstates we obtain a highly collective spin model resulting in dynamics robust to single-particle dephasing, generated  e.g. by  inhomogeneous local fields, and chemical reactions in many cases unavoidable  in experiments. We predict the stabilization of many-body  coherence which allows for the generation of large  spin squeezing. 
By combining long-range dipolar interactions \cite{AlexeyPRA,AlexeyPRL,Review_molecules,KadenPRL,KadenPRL2,Spin_exchange_nature_2013}, tunable via electric fields, and mode space lattices \cite{PhysRevResearch.1.033075, martin2013,Smaleeaax2019}, our proposal mitigates major limitations, such as losses and decoherence, and opens a path for the near term exploration of collective many-body physics in dipolar molecules.

Finally, we discuss how coherent state-transfer between rotational levels allows for the implementation of time-reversal protocols that facilitate the utilization of the quantum advantage of  spin squeezed states, without the need of single-photon detection capabilities. Under  current experimental conditions, the ideal implementation of our protocol, can lead to a metrological gain  $\approx 19$ dB beyond  the standard quantum limit  for systems of 1000 molecules, yielding an electric field sensitivity of $\Delta E \approx \SI{188}{(nV/cm)/\sqrt{Hz}}$ at $E=\SI{1}{kV/cm} $, assuming 10 ms phase accumulation time  \cite{PhysRevLett.121.253401}. This is comparable to state-of-the-art demonstrated electric field  sensitivities in trapped ion crystals \cite{Bollinger_2020}, and  Rydberg setups \cite{Jing2020} and could be  improved with better  electric field stability and rotational state coherence. Beyond electric field sensing, the realisation of a spin squeezed molecular gas could have a major impact on precision measurements  where the specific advantages of molecules for fundamental physics tests can be leveraged in addition to the quantum advantage brought by spin squeezing.

 The proposed protocol not only opens a path towards the use of quantum degenerate molecular  fermionic   gases for enhanced electromagnetic field sensing, but in parallel the ability of time-reversal opens up opportunities to study many-body non-equilibrium dynamics and quantum chaos via out-of-time ordered correlators.

 \begin{acknowledgments}
\noindent{\textit{Acknowledgements:}} 
We thank Diego Barberena and Itamar Kimchi for feedback on the paper and the anonymous referees for providing valuable suggestions.
This work is supported by DARPA and ARO grant W911NF-16-1-0576, W911NF-19-1-0210, NSF PHY1820885, NSF JILA-PFC PHY-1734006 grants, QLCI-2016244, Cold-Molecules AFOSR-MURI, and by NIST.
\end{acknowledgments}


\nocite{Truman1999,Bruun_2012,Schafer_2012,Enss_2012,Baur_2013,Bohn_2009}

%


\begin{thebibliography}{61}%
\makeatletter
\providecommand \@ifxundefined [1]{%
 \@ifx{#1\undefined}
}%
\providecommand \@ifnum [1]{%
 \ifnum #1\expandafter \@firstoftwo
 \else \expandafter \@secondoftwo
 \fi
}%
\providecommand \@ifx [1]{%
 \ifx #1\expandafter \@firstoftwo
 \else \expandafter \@secondoftwo
 \fi
}%
\providecommand \natexlab [1]{#1}%
\providecommand \enquote  [1]{``#1''}%
\providecommand \bibnamefont  [1]{#1}%
\providecommand \bibfnamefont [1]{#1}%
\providecommand \citenamefont [1]{#1}%
\providecommand \href@noop [0]{\@secondoftwo}%
\providecommand \href [0]{\begingroup \@sanitize@url \@href}%
\providecommand \@href[1]{\@@startlink{#1}\@@href}%
\providecommand \@@href[1]{\endgroup#1\@@endlink}%
\providecommand \@sanitize@url [0]{\catcode `\\12\catcode `\$12\catcode
  `\&12\catcode `\#12\catcode `\^12\catcode `\_12\catcode `\%12\relax}%
\providecommand \@@startlink[1]{}%
\providecommand \@@endlink[0]{}%
\providecommand \url  [0]{\begingroup\@sanitize@url \@url }%
\providecommand \@url [1]{\endgroup\@href {#1}{\urlprefix }}%
\providecommand \urlprefix  [0]{URL }%
\providecommand \Eprint [0]{\href }%
\providecommand \doibase [0]{https://doi.org/}%
\providecommand \selectlanguage [0]{\@gobble}%
\providecommand \bibinfo  [0]{\@secondoftwo}%
\providecommand \bibfield  [0]{\@secondoftwo}%
\providecommand \translation [1]{[#1]}%
\providecommand \BibitemOpen [0]{}%
\providecommand \bibitemStop [0]{}%
\providecommand \bibitemNoStop [0]{.\EOS\space}%
\providecommand \EOS [0]{\spacefactor3000\relax}%
\providecommand \BibitemShut  [1]{\csname bibitem#1\endcsname}%
\let\auto@bib@innerbib\@empty
\bibitem [{\citenamefont {Baranov}\ \emph {et~al.}(2012)\citenamefont
  {Baranov}, \citenamefont {Dalmonte}, \citenamefont {Pupillo},\ and\
  \citenamefont {Zoller}}]{Review_dipolar}%
  \BibitemOpen
  \bibfield  {author} {\bibinfo {author} {\bibfnamefont {M.~A.}\ \bibnamefont
  {Baranov}}, \bibinfo {author} {\bibfnamefont {M.}~\bibnamefont {Dalmonte}},
  \bibinfo {author} {\bibfnamefont {G.}~\bibnamefont {Pupillo}},\ and\ \bibinfo
  {author} {\bibfnamefont {P.}~\bibnamefont {Zoller}},\ }\bibfield  {title}
  {\bibinfo {title} {Condensed matter theory of dipolar quantum gases},\ }\href
  {https://doi.org/10.1021/cr2003568} {\bibfield  {journal} {\bibinfo
  {journal} {Chemical Reviews}\ }\textbf {\bibinfo {volume} {112}},\ \bibinfo
  {pages} {5012–5061} (\bibinfo {year} {2012})}\BibitemShut {NoStop}%
\bibitem [{\citenamefont {Bohn}\ \emph {et~al.}(2017)\citenamefont {Bohn},
  \citenamefont {Rey},\ and\ \citenamefont {Ye}}]{Review_molecules}%
  \BibitemOpen
  \bibfield  {author} {\bibinfo {author} {\bibfnamefont {J.~L.}\ \bibnamefont
  {Bohn}}, \bibinfo {author} {\bibfnamefont {A.~M.}\ \bibnamefont {Rey}},\ and\
  \bibinfo {author} {\bibfnamefont {J.}~\bibnamefont {Ye}},\ }\bibfield
  {title} {\bibinfo {title} {Cold molecules: Progress in quantum engineering of
  chemistry and quantum matter},\ }\href
  {https://doi.org/10.1126/science.aam6299} {\bibfield  {journal} {\bibinfo
  {journal} {Science}\ }\textbf {\bibinfo {volume} {357}},\ \bibinfo {pages}
  {1002–1010} (\bibinfo {year} {2017})}\BibitemShut {NoStop}%
\bibitem [{\citenamefont {Micheli}\ \emph {et~al.}(2006)\citenamefont
  {Micheli}, \citenamefont {Brennen},\ and\ \citenamefont
  {Zoller}}]{Micheli2006}%
  \BibitemOpen
  \bibfield  {author} {\bibinfo {author} {\bibfnamefont {A.}~\bibnamefont
  {Micheli}}, \bibinfo {author} {\bibfnamefont {G.~K.}\ \bibnamefont
  {Brennen}},\ and\ \bibinfo {author} {\bibfnamefont {P.}~\bibnamefont
  {Zoller}},\ }\bibfield  {title} {\bibinfo {title} {A toolbox for lattice-spin
  models with polar molecules},\ }\href {https://doi.org/10.1038/nphys287}
  {\bibfield  {journal} {\bibinfo  {journal} {Nature Physics}\ }\textbf
  {\bibinfo {volume} {2}},\ \bibinfo {pages} {341–347} (\bibinfo {year}
  {2006})}\BibitemShut {NoStop}%
\bibitem [{\citenamefont {Barnett}\ \emph {et~al.}(2006)\citenamefont
  {Barnett}, \citenamefont {Petrov}, \citenamefont {Lukin},\ and\ \citenamefont
  {Demler}}]{PhysRevLett.96.190401}%
  \BibitemOpen
  \bibfield  {author} {\bibinfo {author} {\bibfnamefont {R.}~\bibnamefont
  {Barnett}}, \bibinfo {author} {\bibfnamefont {D.}~\bibnamefont {Petrov}},
  \bibinfo {author} {\bibfnamefont {M.}~\bibnamefont {Lukin}},\ and\ \bibinfo
  {author} {\bibfnamefont {E.}~\bibnamefont {Demler}},\ }\bibfield  {title}
  {\bibinfo {title} {Quantum magnetism with multicomponent dipolar molecules in
  an optical lattice},\ }\href {https://doi.org/10.1103/PhysRevLett.96.190401}
  {\bibfield  {journal} {\bibinfo  {journal} {Phys. Rev. Lett.}\ }\textbf
  {\bibinfo {volume} {96}},\ \bibinfo {pages} {190401} (\bibinfo {year}
  {2006})}\BibitemShut {NoStop}%
\bibitem [{\citenamefont {Gorshkov}\ \emph
  {et~al.}(2011{\natexlab{a}})\citenamefont {Gorshkov}, \citenamefont
  {Manmana}, \citenamefont {Chen}, \citenamefont {Demler}, \citenamefont
  {Lukin},\ and\ \citenamefont {Rey}}]{AlexeyPRA}%
  \BibitemOpen
  \bibfield  {author} {\bibinfo {author} {\bibfnamefont {A.~V.}\ \bibnamefont
  {Gorshkov}}, \bibinfo {author} {\bibfnamefont {S.~R.}\ \bibnamefont
  {Manmana}}, \bibinfo {author} {\bibfnamefont {G.}~\bibnamefont {Chen}},
  \bibinfo {author} {\bibfnamefont {E.}~\bibnamefont {Demler}}, \bibinfo
  {author} {\bibfnamefont {M.~D.}\ \bibnamefont {Lukin}},\ and\ \bibinfo
  {author} {\bibfnamefont {A.~M.}\ \bibnamefont {Rey}},\ }\bibfield  {title}
  {\bibinfo {title} {Quantum magnetism with polar alkali-metal dimers},\ }\href
  {https://doi.org/10.1103/PhysRevA.84.033619} {\bibfield  {journal} {\bibinfo
  {journal} {Phys. Rev. A}\ }\textbf {\bibinfo {volume} {84}},\ \bibinfo
  {pages} {033619} (\bibinfo {year} {2011}{\natexlab{a}})}\BibitemShut
  {NoStop}%
\bibitem [{\citenamefont {Gorshkov}\ \emph
  {et~al.}(2011{\natexlab{b}})\citenamefont {Gorshkov}, \citenamefont
  {Manmana}, \citenamefont {Chen}, \citenamefont {Ye}, \citenamefont {Demler},
  \citenamefont {Lukin},\ and\ \citenamefont {Rey}}]{AlexeyPRL}%
  \BibitemOpen
  \bibfield  {author} {\bibinfo {author} {\bibfnamefont {A.~V.}\ \bibnamefont
  {Gorshkov}}, \bibinfo {author} {\bibfnamefont {S.~R.}\ \bibnamefont
  {Manmana}}, \bibinfo {author} {\bibfnamefont {G.}~\bibnamefont {Chen}},
  \bibinfo {author} {\bibfnamefont {J.}~\bibnamefont {Ye}}, \bibinfo {author}
  {\bibfnamefont {E.}~\bibnamefont {Demler}}, \bibinfo {author} {\bibfnamefont
  {M.~D.}\ \bibnamefont {Lukin}},\ and\ \bibinfo {author} {\bibfnamefont
  {A.~M.}\ \bibnamefont {Rey}},\ }\bibfield  {title} {\bibinfo {title} {Tunable
  superfluidity and quantum magnetism with ultracold polar molecules},\ }\href
  {https://doi.org/10.1103/PhysRevLett.107.115301} {\bibfield  {journal}
  {\bibinfo  {journal} {Phys. Rev. Lett.}\ }\textbf {\bibinfo {volume} {107}},\
  \bibinfo {pages} {115301} (\bibinfo {year} {2011}{\natexlab{b}})}\BibitemShut
  {NoStop}%
\bibitem [{\citenamefont {Hazzard}\ \emph {et~al.}(2013)\citenamefont
  {Hazzard}, \citenamefont {Manmana}, \citenamefont {Foss-Feig},\ and\
  \citenamefont {Rey}}]{KadenPRL}%
  \BibitemOpen
  \bibfield  {author} {\bibinfo {author} {\bibfnamefont {K.~R.~A.}\
  \bibnamefont {Hazzard}}, \bibinfo {author} {\bibfnamefont {S.~R.}\
  \bibnamefont {Manmana}}, \bibinfo {author} {\bibfnamefont {M.}~\bibnamefont
  {Foss-Feig}},\ and\ \bibinfo {author} {\bibfnamefont {A.~M.}\ \bibnamefont
  {Rey}},\ }\bibfield  {title} {\bibinfo {title} {Far-from-equilibrium quantum
  magnetism with ultracold polar molecules},\ }\href
  {https://doi.org/10.1103/PhysRevLett.110.075301} {\bibfield  {journal}
  {\bibinfo  {journal} {Phys. Rev. Lett.}\ }\textbf {\bibinfo {volume} {110}},\
  \bibinfo {pages} {075301} (\bibinfo {year} {2013})}\BibitemShut {NoStop}%
\bibitem [{\citenamefont {Hazzard}\ \emph {et~al.}(2014)\citenamefont
  {Hazzard}, \citenamefont {Gadway}, \citenamefont {Foss-Feig}, \citenamefont
  {Yan}, \citenamefont {Moses}, \citenamefont {Covey}, \citenamefont {Yao},
  \citenamefont {Lukin}, \citenamefont {Ye}, \citenamefont {Jin},\ and\
  \citenamefont {Rey}}]{KadenPRL2}%
  \BibitemOpen
  \bibfield  {author} {\bibinfo {author} {\bibfnamefont {K.~R.~A.}\
  \bibnamefont {Hazzard}}, \bibinfo {author} {\bibfnamefont {B.}~\bibnamefont
  {Gadway}}, \bibinfo {author} {\bibfnamefont {M.}~\bibnamefont {Foss-Feig}},
  \bibinfo {author} {\bibfnamefont {B.}~\bibnamefont {Yan}}, \bibinfo {author}
  {\bibfnamefont {S.~A.}\ \bibnamefont {Moses}}, \bibinfo {author}
  {\bibfnamefont {J.~P.}\ \bibnamefont {Covey}}, \bibinfo {author}
  {\bibfnamefont {N.~Y.}\ \bibnamefont {Yao}}, \bibinfo {author} {\bibfnamefont
  {M.~D.}\ \bibnamefont {Lukin}}, \bibinfo {author} {\bibfnamefont
  {J.}~\bibnamefont {Ye}}, \bibinfo {author} {\bibfnamefont {D.~S.}\
  \bibnamefont {Jin}},\ and\ \bibinfo {author} {\bibfnamefont {A.~M.}\
  \bibnamefont {Rey}},\ }\bibfield  {title} {\bibinfo {title} {Many-body
  dynamics of dipolar molecules in an optical lattice},\ }\href
  {https://doi.org/10.1103/PhysRevLett.113.195302} {\bibfield  {journal}
  {\bibinfo  {journal} {Phys. Rev. Lett.}\ }\textbf {\bibinfo {volume} {113}},\
  \bibinfo {pages} {195302} (\bibinfo {year} {2014})}\BibitemShut {NoStop}%
\bibitem [{\citenamefont {DeMille}\ \emph {et~al.}(2017)\citenamefont
  {DeMille}, \citenamefont {Doyle},\ and\ \citenamefont
  {Sushkov}}]{DeMille990}%
  \BibitemOpen
  \bibfield  {author} {\bibinfo {author} {\bibfnamefont {D.}~\bibnamefont
  {DeMille}}, \bibinfo {author} {\bibfnamefont {J.~M.}\ \bibnamefont {Doyle}},\
  and\ \bibinfo {author} {\bibfnamefont {A.~O.}\ \bibnamefont {Sushkov}},\
  }\bibfield  {title} {\bibinfo {title} {Probing the frontiers of particle
  physics with tabletop-scale experiments},\ }\href
  {https://doi.org/10.1126/science.aal3003} {\bibfield  {journal} {\bibinfo
  {journal} {Science}\ }\textbf {\bibinfo {volume} {357}},\ \bibinfo {pages}
  {990–994} (\bibinfo {year} {2017})}\BibitemShut {NoStop}%
\bibitem [{\citenamefont {André}\ \emph {et~al.}(2006)\citenamefont {André},
  \citenamefont {DeMille}, \citenamefont {Doyle}, \citenamefont {Lukin},
  \citenamefont {Maxwell}, \citenamefont {Rabl}, \citenamefont {Schoelkopf},\
  and\ \citenamefont {Zoller}}]{Andre2006}%
  \BibitemOpen
  \bibfield  {author} {\bibinfo {author} {\bibfnamefont {A.}~\bibnamefont
  {André}}, \bibinfo {author} {\bibfnamefont {D.}~\bibnamefont {DeMille}},
  \bibinfo {author} {\bibfnamefont {J.~M.}\ \bibnamefont {Doyle}}, \bibinfo
  {author} {\bibfnamefont {M.~D.}\ \bibnamefont {Lukin}}, \bibinfo {author}
  {\bibfnamefont {S.~E.}\ \bibnamefont {Maxwell}}, \bibinfo {author}
  {\bibfnamefont {P.}~\bibnamefont {Rabl}}, \bibinfo {author} {\bibfnamefont
  {R.~J.}\ \bibnamefont {Schoelkopf}},\ and\ \bibinfo {author} {\bibfnamefont
  {P.}~\bibnamefont {Zoller}},\ }\bibfield  {title} {\bibinfo {title} {A
  coherent all-electrical interface between polar molecules and mesoscopic
  superconducting resonators},\ }\href {https://doi.org/10.1038/nphys386}
  {\bibfield  {journal} {\bibinfo  {journal} {Nature Physics}\ }\textbf
  {\bibinfo {volume} {2}},\ \bibinfo {pages} {636–642} (\bibinfo {year}
  {2006})}\BibitemShut {NoStop}%
\bibitem [{\citenamefont {Pezz\`e}\ \emph {et~al.}(2018)\citenamefont
  {Pezz\`e}, \citenamefont {Smerzi}, \citenamefont {Oberthaler}, \citenamefont
  {Schmied},\ and\ \citenamefont {Treutlein}}]{RevMod_Metrology_2018}%
  \BibitemOpen
  \bibfield  {author} {\bibinfo {author} {\bibfnamefont {L.}~\bibnamefont
  {Pezz\`e}}, \bibinfo {author} {\bibfnamefont {A.}~\bibnamefont {Smerzi}},
  \bibinfo {author} {\bibfnamefont {M.~K.}\ \bibnamefont {Oberthaler}},
  \bibinfo {author} {\bibfnamefont {R.}~\bibnamefont {Schmied}},\ and\ \bibinfo
  {author} {\bibfnamefont {P.}~\bibnamefont {Treutlein}},\ }\bibfield  {title}
  {\bibinfo {title} {Quantum metrology with nonclassical states of atomic
  ensembles},\ }\href {https://doi.org/10.1103/RevModPhys.90.035005} {\bibfield
   {journal} {\bibinfo  {journal} {Rev. Mod. Phys.}\ }\textbf {\bibinfo
  {volume} {90}},\ \bibinfo {pages} {035005} (\bibinfo {year}
  {2018})}\BibitemShut {NoStop}%
\bibitem [{\citenamefont {Ni}\ \emph {et~al.}(2010)\citenamefont {Ni},
  \citenamefont {Ospelkaus}, \citenamefont {Wang}, \citenamefont {Quéméner},
  \citenamefont {Neyenhuis}, \citenamefont {de~Miranda}, \citenamefont {Bohn},
  \citenamefont {Ye},\ and\ \citenamefont {Jin}}]{Ni2010}%
  \BibitemOpen
  \bibfield  {author} {\bibinfo {author} {\bibfnamefont {K.-K.}\ \bibnamefont
  {Ni}}, \bibinfo {author} {\bibfnamefont {S.}~\bibnamefont {Ospelkaus}},
  \bibinfo {author} {\bibfnamefont {D.}~\bibnamefont {Wang}}, \bibinfo {author}
  {\bibfnamefont {G.}~\bibnamefont {Quéméner}}, \bibinfo {author}
  {\bibfnamefont {B.}~\bibnamefont {Neyenhuis}}, \bibinfo {author}
  {\bibfnamefont {M.~H.~G.}\ \bibnamefont {de~Miranda}}, \bibinfo {author}
  {\bibfnamefont {J.~L.}\ \bibnamefont {Bohn}}, \bibinfo {author}
  {\bibfnamefont {J.}~\bibnamefont {Ye}},\ and\ \bibinfo {author}
  {\bibfnamefont {D.~S.}\ \bibnamefont {Jin}},\ }\bibfield  {title} {\bibinfo
  {title} {Dipolar collisions of polar molecules in the quantum regime},\
  }\href {https://doi.org/10.1038/nature08953} {\bibfield  {journal} {\bibinfo
  {journal} {Nature}\ }\textbf {\bibinfo {volume} {464}},\ \bibinfo {pages}
  {1324–1328} (\bibinfo {year} {2010})}\BibitemShut {NoStop}%
\bibitem [{\citenamefont {Guo}\ \emph {et~al.}(2018)\citenamefont {Guo},
  \citenamefont {Ye}, \citenamefont {He}, \citenamefont
  {Gonz\'alez-Mart\'{\i}nez}, \citenamefont {Vexiau}, \citenamefont
  {Qu\'em\'ener},\ and\ \citenamefont {Wang}}]{PhysRevX.8.041044}%
  \BibitemOpen
  \bibfield  {author} {\bibinfo {author} {\bibfnamefont {M.}~\bibnamefont
  {Guo}}, \bibinfo {author} {\bibfnamefont {X.}~\bibnamefont {Ye}}, \bibinfo
  {author} {\bibfnamefont {J.}~\bibnamefont {He}}, \bibinfo {author}
  {\bibfnamefont {M.~L.}\ \bibnamefont {Gonz\'alez-Mart\'{\i}nez}}, \bibinfo
  {author} {\bibfnamefont {R.}~\bibnamefont {Vexiau}}, \bibinfo {author}
  {\bibfnamefont {G.}~\bibnamefont {Qu\'em\'ener}},\ and\ \bibinfo {author}
  {\bibfnamefont {D.}~\bibnamefont {Wang}},\ }\bibfield  {title} {\bibinfo
  {title} {Dipolar collisions of ultracold ground-state bosonic molecules},\
  }\href {https://doi.org/10.1103/PhysRevX.8.041044} {\bibfield  {journal}
  {\bibinfo  {journal} {Phys. Rev. X}\ }\textbf {\bibinfo {volume} {8}},\
  \bibinfo {pages} {041044} (\bibinfo {year} {2018})}\BibitemShut {NoStop}%
\bibitem [{\citenamefont {Ospelkaus}\ \emph {et~al.}(2010)\citenamefont
  {Ospelkaus}, \citenamefont {Ni}, \citenamefont {Wang}, \citenamefont
  {de~Miranda}, \citenamefont {Neyenhuis}, \citenamefont {Quéméner},
  \citenamefont {Julienne}, \citenamefont {Bohn}, \citenamefont {Jin},\ and\
  \citenamefont {Ye}}]{Ospelkaus853}%
  \BibitemOpen
  \bibfield  {author} {\bibinfo {author} {\bibfnamefont {S.}~\bibnamefont
  {Ospelkaus}}, \bibinfo {author} {\bibfnamefont {K.-K.}\ \bibnamefont {Ni}},
  \bibinfo {author} {\bibfnamefont {D.}~\bibnamefont {Wang}}, \bibinfo {author}
  {\bibfnamefont {M.~H.~G.}\ \bibnamefont {de~Miranda}}, \bibinfo {author}
  {\bibfnamefont {B.}~\bibnamefont {Neyenhuis}}, \bibinfo {author}
  {\bibfnamefont {G.}~\bibnamefont {Quéméner}}, \bibinfo {author}
  {\bibfnamefont {P.~S.}\ \bibnamefont {Julienne}}, \bibinfo {author}
  {\bibfnamefont {J.~L.}\ \bibnamefont {Bohn}}, \bibinfo {author}
  {\bibfnamefont {D.~S.}\ \bibnamefont {Jin}},\ and\ \bibinfo {author}
  {\bibfnamefont {J.}~\bibnamefont {Ye}},\ }\bibfield  {title} {\bibinfo
  {title} {Quantum-state controlled chemical reactions of ultracold
  potassium-rubidium molecules},\ }\href
  {https://doi.org/10.1126/science.1184121} {\bibfield  {journal} {\bibinfo
  {journal} {Science}\ }\textbf {\bibinfo {volume} {327}},\ \bibinfo {pages}
  {853–857} (\bibinfo {year} {2010})}\BibitemShut {NoStop}%
\bibitem [{\citenamefont {Gregory}\ \emph {et~al.}(2019)\citenamefont
  {Gregory}, \citenamefont {Frye}, \citenamefont {Blackmore}, \citenamefont
  {Bridge}, \citenamefont {Sawant}, \citenamefont {Hutson},\ and\ \citenamefont
  {Cornish}}]{Gregory2019}%
  \BibitemOpen
  \bibfield  {author} {\bibinfo {author} {\bibfnamefont {P.~D.}\ \bibnamefont
  {Gregory}}, \bibinfo {author} {\bibfnamefont {M.~D.}\ \bibnamefont {Frye}},
  \bibinfo {author} {\bibfnamefont {J.~A.}\ \bibnamefont {Blackmore}}, \bibinfo
  {author} {\bibfnamefont {E.~M.}\ \bibnamefont {Bridge}}, \bibinfo {author}
  {\bibfnamefont {R.}~\bibnamefont {Sawant}}, \bibinfo {author} {\bibfnamefont
  {J.~M.}\ \bibnamefont {Hutson}},\ and\ \bibinfo {author} {\bibfnamefont
  {S.~L.}\ \bibnamefont {Cornish}},\ }\bibfield  {title} {\bibinfo {title}
  {Sticky collisions of ultracold rbcs molecules},\ }\href
  {https://doi.org/10.1038/s41467-019-11033-y} {\bibfield  {journal} {\bibinfo
  {journal} {Nature Communications}\ }\textbf {\bibinfo {volume} {10}},\
  \bibinfo {pages} {3104} (\bibinfo {year} {2019})}\BibitemShut {NoStop}%
\bibitem [{\citenamefont {B\"uchler}\ \emph {et~al.}(2007)\citenamefont
  {B\"uchler}, \citenamefont {Demler}, \citenamefont {Lukin}, \citenamefont
  {Micheli}, \citenamefont {Prokof'ev}, \citenamefont {Pupillo},\ and\
  \citenamefont {Zoller}}]{PhysRevLett.98.060404}%
  \BibitemOpen
  \bibfield  {author} {\bibinfo {author} {\bibfnamefont {H.~P.}\ \bibnamefont
  {B\"uchler}}, \bibinfo {author} {\bibfnamefont {E.}~\bibnamefont {Demler}},
  \bibinfo {author} {\bibfnamefont {M.}~\bibnamefont {Lukin}}, \bibinfo
  {author} {\bibfnamefont {A.}~\bibnamefont {Micheli}}, \bibinfo {author}
  {\bibfnamefont {N.}~\bibnamefont {Prokof'ev}}, \bibinfo {author}
  {\bibfnamefont {G.}~\bibnamefont {Pupillo}},\ and\ \bibinfo {author}
  {\bibfnamefont {P.}~\bibnamefont {Zoller}},\ }\bibfield  {title} {\bibinfo
  {title} {Strongly correlated 2d quantum phases with cold polar molecules:
  Controlling the shape of the interaction potential},\ }\href
  {https://doi.org/10.1103/PhysRevLett.98.060404} {\bibfield  {journal}
  {\bibinfo  {journal} {Phys. Rev. Lett.}\ }\textbf {\bibinfo {volume} {98}},\
  \bibinfo {pages} {060404} (\bibinfo {year} {2007})}\BibitemShut {NoStop}%
\bibitem [{\citenamefont {de~Miranda}\ \emph {et~al.}(2011)\citenamefont
  {de~Miranda}, \citenamefont {Chotia}, \citenamefont {Neyenhuis},
  \citenamefont {Wang}, \citenamefont {Quéméner}, \citenamefont {Ospelkaus},
  \citenamefont {Bohn}, \citenamefont {Ye},\ and\ \citenamefont
  {Jin}}]{deMiranda2011}%
  \BibitemOpen
  \bibfield  {author} {\bibinfo {author} {\bibfnamefont {M.~H.~G.}\
  \bibnamefont {de~Miranda}}, \bibinfo {author} {\bibfnamefont
  {A.}~\bibnamefont {Chotia}}, \bibinfo {author} {\bibfnamefont
  {B.}~\bibnamefont {Neyenhuis}}, \bibinfo {author} {\bibfnamefont
  {D.}~\bibnamefont {Wang}}, \bibinfo {author} {\bibfnamefont {G.}~\bibnamefont
  {Quéméner}}, \bibinfo {author} {\bibfnamefont {S.}~\bibnamefont
  {Ospelkaus}}, \bibinfo {author} {\bibfnamefont {J.~L.}\ \bibnamefont {Bohn}},
  \bibinfo {author} {\bibfnamefont {J.}~\bibnamefont {Ye}},\ and\ \bibinfo
  {author} {\bibfnamefont {D.~S.}\ \bibnamefont {Jin}},\ }\bibfield  {title}
  {\bibinfo {title} {Controlling the quantum stereodynamics of ultracold
  bimolecular reactions},\ }\href {https://doi.org/10.1038/nphys1939}
  {\bibfield  {journal} {\bibinfo  {journal} {Nature Physics}\ }\textbf
  {\bibinfo {volume} {7}},\ \bibinfo {pages} {502–507} (\bibinfo {year}
  {2011})}\BibitemShut {NoStop}%
\bibitem [{\citenamefont {Chotia}\ \emph {et~al.}(2012)\citenamefont {Chotia},
  \citenamefont {Neyenhuis}, \citenamefont {Moses}, \citenamefont {Yan},
  \citenamefont {Covey}, \citenamefont {Foss-Feig}, \citenamefont {Rey},
  \citenamefont {Jin},\ and\ \citenamefont {Ye}}]{PhysRevLett.108.080405}%
  \BibitemOpen
  \bibfield  {author} {\bibinfo {author} {\bibfnamefont {A.}~\bibnamefont
  {Chotia}}, \bibinfo {author} {\bibfnamefont {B.}~\bibnamefont {Neyenhuis}},
  \bibinfo {author} {\bibfnamefont {S.~A.}\ \bibnamefont {Moses}}, \bibinfo
  {author} {\bibfnamefont {B.}~\bibnamefont {Yan}}, \bibinfo {author}
  {\bibfnamefont {J.~P.}\ \bibnamefont {Covey}}, \bibinfo {author}
  {\bibfnamefont {M.}~\bibnamefont {Foss-Feig}}, \bibinfo {author}
  {\bibfnamefont {A.~M.}\ \bibnamefont {Rey}}, \bibinfo {author} {\bibfnamefont
  {D.~S.}\ \bibnamefont {Jin}},\ and\ \bibinfo {author} {\bibfnamefont
  {J.}~\bibnamefont {Ye}},\ }\bibfield  {title} {\bibinfo {title} {Long-lived
  dipolar molecules and feshbach molecules in a 3d optical lattice},\ }\href
  {https://doi.org/10.1103/PhysRevLett.108.080405} {\bibfield  {journal}
  {\bibinfo  {journal} {Phys. Rev. Lett.}\ }\textbf {\bibinfo {volume} {108}},\
  \bibinfo {pages} {080405} (\bibinfo {year} {2012})}\BibitemShut {NoStop}%
\bibitem [{\citenamefont {Moses}\ \emph {et~al.}(2015)\citenamefont {Moses},
  \citenamefont {Covey}, \citenamefont {Miecnikowski}, \citenamefont {Yan},
  \citenamefont {Gadway}, \citenamefont {Ye},\ and\ \citenamefont
  {Jin}}]{Moses659}%
  \BibitemOpen
  \bibfield  {author} {\bibinfo {author} {\bibfnamefont {S.~A.}\ \bibnamefont
  {Moses}}, \bibinfo {author} {\bibfnamefont {J.~P.}\ \bibnamefont {Covey}},
  \bibinfo {author} {\bibfnamefont {M.~T.}\ \bibnamefont {Miecnikowski}},
  \bibinfo {author} {\bibfnamefont {B.}~\bibnamefont {Yan}}, \bibinfo {author}
  {\bibfnamefont {B.}~\bibnamefont {Gadway}}, \bibinfo {author} {\bibfnamefont
  {J.}~\bibnamefont {Ye}},\ and\ \bibinfo {author} {\bibfnamefont {D.~S.}\
  \bibnamefont {Jin}},\ }\bibfield  {title} {\bibinfo {title} {Creation of a
  low-entropy quantum gas of polar molecules in an optical lattice},\ }\href
  {https://doi.org/10.1126/science.aac6400} {\bibfield  {journal} {\bibinfo
  {journal} {Science}\ }\textbf {\bibinfo {volume} {350}},\ \bibinfo {pages}
  {659–662} (\bibinfo {year} {2015})}\BibitemShut {NoStop}%
\bibitem [{\citenamefont {Yan}\ \emph {et~al.}(2013)\citenamefont {Yan},
  \citenamefont {Moses}, \citenamefont {Gadway}, \citenamefont {Covey},
  \citenamefont {Hazzard}, \citenamefont {Rey}, \citenamefont {Jin},\ and\
  \citenamefont {Ye}}]{Spin_exchange_nature_2013}%
  \BibitemOpen
  \bibfield  {author} {\bibinfo {author} {\bibfnamefont {B.}~\bibnamefont
  {Yan}}, \bibinfo {author} {\bibfnamefont {S.~A.}\ \bibnamefont {Moses}},
  \bibinfo {author} {\bibfnamefont {B.}~\bibnamefont {Gadway}}, \bibinfo
  {author} {\bibfnamefont {J.~P.}\ \bibnamefont {Covey}}, \bibinfo {author}
  {\bibfnamefont {K.~R.~A.}\ \bibnamefont {Hazzard}}, \bibinfo {author}
  {\bibfnamefont {A.~M.}\ \bibnamefont {Rey}}, \bibinfo {author} {\bibfnamefont
  {D.~S.}\ \bibnamefont {Jin}},\ and\ \bibinfo {author} {\bibfnamefont
  {J.}~\bibnamefont {Ye}},\ }\bibfield  {title} {\bibinfo {title} {Observation
  of dipolar spin-exchange interactions with lattice-confined polar
  molecules},\ }\href {https://doi.org/10.1038/nature12483} {\bibfield
  {journal} {\bibinfo  {journal} {Nature}\ }\textbf {\bibinfo {volume} {501}},\
  \bibinfo {pages} {521–525} (\bibinfo {year} {2013})}\BibitemShut {NoStop}%
\bibitem [{\citenamefont {Zhu}\ \emph {et~al.}(2014)\citenamefont {Zhu},
  \citenamefont {Gadway}, \citenamefont {Foss-Feig}, \citenamefont
  {Schachenmayer}, \citenamefont {Wall}, \citenamefont {Hazzard}, \citenamefont
  {Yan}, \citenamefont {Moses}, \citenamefont {Covey}, \citenamefont {Jin},
  \citenamefont {Ye}, \citenamefont {Holland},\ and\ \citenamefont
  {Rey}}]{Zhu2014}%
  \BibitemOpen
  \bibfield  {author} {\bibinfo {author} {\bibfnamefont {B.}~\bibnamefont
  {Zhu}}, \bibinfo {author} {\bibfnamefont {B.}~\bibnamefont {Gadway}},
  \bibinfo {author} {\bibfnamefont {M.}~\bibnamefont {Foss-Feig}}, \bibinfo
  {author} {\bibfnamefont {J.}~\bibnamefont {Schachenmayer}}, \bibinfo {author}
  {\bibfnamefont {M.~L.}\ \bibnamefont {Wall}}, \bibinfo {author}
  {\bibfnamefont {K.~R.~A.}\ \bibnamefont {Hazzard}}, \bibinfo {author}
  {\bibfnamefont {B.}~\bibnamefont {Yan}}, \bibinfo {author} {\bibfnamefont
  {S.~A.}\ \bibnamefont {Moses}}, \bibinfo {author} {\bibfnamefont {J.~P.}\
  \bibnamefont {Covey}}, \bibinfo {author} {\bibfnamefont {D.~S.}\ \bibnamefont
  {Jin}}, \bibinfo {author} {\bibfnamefont {J.}~\bibnamefont {Ye}}, \bibinfo
  {author} {\bibfnamefont {M.}~\bibnamefont {Holland}},\ and\ \bibinfo {author}
  {\bibfnamefont {A.~M.}\ \bibnamefont {Rey}},\ }\bibfield  {title} {\bibinfo
  {title} {Suppressing the loss of ultracold molecules via the continuous
  quantum zeno effect},\ }\href
  {https://doi.org/10.1103/PhysRevLett.112.070404} {\bibfield  {journal}
  {\bibinfo  {journal} {Phys. Rev. Lett.}\ }\textbf {\bibinfo {volume} {112}},\
  \bibinfo {pages} {070404} (\bibinfo {year} {2014})}\BibitemShut {NoStop}%
\bibitem [{\citenamefont {{De Marco}}\ \emph {et~al.}(2019)\citenamefont {{De
  Marco}}, \citenamefont {Valtolina}, \citenamefont {Matsuda}, \citenamefont
  {Tobias}, \citenamefont {Covey},\ and\ \citenamefont {Ye}}]{DeMarco2019}%
  \BibitemOpen
  \bibfield  {author} {\bibinfo {author} {\bibfnamefont {L.}~\bibnamefont {{De
  Marco}}}, \bibinfo {author} {\bibfnamefont {G.}~\bibnamefont {Valtolina}},
  \bibinfo {author} {\bibfnamefont {K.}~\bibnamefont {Matsuda}}, \bibinfo
  {author} {\bibfnamefont {W.~G.}\ \bibnamefont {Tobias}}, \bibinfo {author}
  {\bibfnamefont {J.~P.}\ \bibnamefont {Covey}},\ and\ \bibinfo {author}
  {\bibfnamefont {J.}~\bibnamefont {Ye}},\ }\bibfield  {title} {\bibinfo
  {title} {A degenerate fermi gas of polar molecules},\ }\href
  {https://doi.org/10.1126/science.aau7230} {\bibfield  {journal} {\bibinfo
  {journal} {Science}\ }\textbf {\bibinfo {volume} {363}},\ \bibinfo {pages}
  {853–856} (\bibinfo {year} {2019})}\BibitemShut {NoStop}%
\bibitem [{\citenamefont {Tobias}\ \emph {et~al.}(2020)\citenamefont {Tobias},
  \citenamefont {Matsuda}, \citenamefont {Valtolina}, \citenamefont {De~Marco},
  \citenamefont {Li},\ and\ \citenamefont {Ye}}]{PhysRevLett.124.033401}%
  \BibitemOpen
  \bibfield  {author} {\bibinfo {author} {\bibfnamefont {W.~G.}\ \bibnamefont
  {Tobias}}, \bibinfo {author} {\bibfnamefont {K.}~\bibnamefont {Matsuda}},
  \bibinfo {author} {\bibfnamefont {G.}~\bibnamefont {Valtolina}}, \bibinfo
  {author} {\bibfnamefont {L.}~\bibnamefont {De~Marco}}, \bibinfo {author}
  {\bibfnamefont {J.-R.}\ \bibnamefont {Li}},\ and\ \bibinfo {author}
  {\bibfnamefont {J.}~\bibnamefont {Ye}},\ }\bibfield  {title} {\bibinfo
  {title} {Thermalization and sub-poissonian density fluctuations in a
  degenerate molecular fermi gas},\ }\href
  {https://doi.org/10.1103/PhysRevLett.124.033401} {\bibfield  {journal}
  {\bibinfo  {journal} {Phys. Rev. Lett.}\ }\textbf {\bibinfo {volume} {124}},\
  \bibinfo {pages} {033401} (\bibinfo {year} {2020})}\BibitemShut {NoStop}%
\bibitem [{\citenamefont {Gregory}\ \emph {et~al.}(2020)\citenamefont
  {Gregory}, \citenamefont {Blackmore}, \citenamefont {Bromley},\ and\
  \citenamefont {Cornish}}]{PhysRevLett.124.163402}%
  \BibitemOpen
  \bibfield  {author} {\bibinfo {author} {\bibfnamefont {P.~D.}\ \bibnamefont
  {Gregory}}, \bibinfo {author} {\bibfnamefont {J.~A.}\ \bibnamefont
  {Blackmore}}, \bibinfo {author} {\bibfnamefont {S.~L.}\ \bibnamefont
  {Bromley}},\ and\ \bibinfo {author} {\bibfnamefont {S.~L.}\ \bibnamefont
  {Cornish}},\ }\bibfield  {title} {\bibinfo {title} {Loss of ultracold
  $^{87}\mathrm{Rb}^{133}\mathrm{Cs}$ molecules via optical excitation of
  long-lived two-body collision complexes},\ }\href
  {https://doi.org/10.1103/PhysRevLett.124.163402} {\bibfield  {journal}
  {\bibinfo  {journal} {Phys. Rev. Lett.}\ }\textbf {\bibinfo {volume} {124}},\
  \bibinfo {pages} {163402} (\bibinfo {year} {2020})}\BibitemShut {NoStop}%
\bibitem [{\citenamefont {Hu}\ \emph {et~al.}(2019)\citenamefont {Hu},
  \citenamefont {Liu}, \citenamefont {Grimes}, \citenamefont {Lin},
  \citenamefont {Gheorghe}, \citenamefont {Vexiau}, \citenamefont
  {Bouloufa-Maafa}, \citenamefont {Dulieu}, \citenamefont {Rosenband},\ and\
  \citenamefont {Ni}}]{Hu1111}%
  \BibitemOpen
  \bibfield  {author} {\bibinfo {author} {\bibfnamefont {M.-G.}\ \bibnamefont
  {Hu}}, \bibinfo {author} {\bibfnamefont {Y.}~\bibnamefont {Liu}}, \bibinfo
  {author} {\bibfnamefont {D.~D.}\ \bibnamefont {Grimes}}, \bibinfo {author}
  {\bibfnamefont {Y.-W.}\ \bibnamefont {Lin}}, \bibinfo {author} {\bibfnamefont
  {A.~H.}\ \bibnamefont {Gheorghe}}, \bibinfo {author} {\bibfnamefont
  {R.}~\bibnamefont {Vexiau}}, \bibinfo {author} {\bibfnamefont
  {N.}~\bibnamefont {Bouloufa-Maafa}}, \bibinfo {author} {\bibfnamefont
  {O.}~\bibnamefont {Dulieu}}, \bibinfo {author} {\bibfnamefont
  {T.}~\bibnamefont {Rosenband}},\ and\ \bibinfo {author} {\bibfnamefont
  {K.-K.}\ \bibnamefont {Ni}},\ }\bibfield  {title} {\bibinfo {title} {Direct
  observation of bimolecular reactions of ultracold krb molecules},\ }\href
  {https://doi.org/10.1126/science.aay9531} {\bibfield  {journal} {\bibinfo
  {journal} {Science}\ }\textbf {\bibinfo {volume} {366}},\ \bibinfo {pages}
  {1111–1115} (\bibinfo {year} {2019})}\BibitemShut {NoStop}%
\bibitem [{\citenamefont {Kirste}\ \emph {et~al.}(2012)\citenamefont {Kirste},
  \citenamefont {Wang}, \citenamefont {Schewe}, \citenamefont {Meijer},
  \citenamefont {Liu}, \citenamefont {van~der Avoird}, \citenamefont {Janssen},
  \citenamefont {Gubbels}, \citenamefont {Groenenboom},\ and\ \citenamefont
  {van~de Meerakker}}]{Kirste1060}%
  \BibitemOpen
  \bibfield  {author} {\bibinfo {author} {\bibfnamefont {M.}~\bibnamefont
  {Kirste}}, \bibinfo {author} {\bibfnamefont {X.}~\bibnamefont {Wang}},
  \bibinfo {author} {\bibfnamefont {H.~C.}\ \bibnamefont {Schewe}}, \bibinfo
  {author} {\bibfnamefont {G.}~\bibnamefont {Meijer}}, \bibinfo {author}
  {\bibfnamefont {K.}~\bibnamefont {Liu}}, \bibinfo {author} {\bibfnamefont
  {A.}~\bibnamefont {van~der Avoird}}, \bibinfo {author} {\bibfnamefont
  {L.~M.~C.}\ \bibnamefont {Janssen}}, \bibinfo {author} {\bibfnamefont
  {K.~B.}\ \bibnamefont {Gubbels}}, \bibinfo {author} {\bibfnamefont {G.~C.}\
  \bibnamefont {Groenenboom}},\ and\ \bibinfo {author} {\bibfnamefont
  {S.~Y.~T.}\ \bibnamefont {van~de Meerakker}},\ }\bibfield  {title} {\bibinfo
  {title} {Quantum-state resolved bimolecular collisions of velocity-controlled
  oh with no radicals},\ }\href {https://doi.org/10.1126/science.1229549}
  {\bibfield  {journal} {\bibinfo  {journal} {Science}\ }\textbf {\bibinfo
  {volume} {338}},\ \bibinfo {pages} {1060–1063} (\bibinfo {year}
  {2012})}\BibitemShut {NoStop}%
\bibitem [{\citenamefont {Christianen}\ \emph {et~al.}(2019)\citenamefont
  {Christianen}, \citenamefont {Zwierlein}, \citenamefont {Groenenboom},\ and\
  \citenamefont {Karman}}]{PhysRevLett.123.123402}%
  \BibitemOpen
  \bibfield  {author} {\bibinfo {author} {\bibfnamefont {A.}~\bibnamefont
  {Christianen}}, \bibinfo {author} {\bibfnamefont {M.~W.}\ \bibnamefont
  {Zwierlein}}, \bibinfo {author} {\bibfnamefont {G.~C.}\ \bibnamefont
  {Groenenboom}},\ and\ \bibinfo {author} {\bibfnamefont {T.}~\bibnamefont
  {Karman}},\ }\bibfield  {title} {\bibinfo {title} {Photoinduced two-body loss
  of ultracold molecules},\ }\href
  {https://doi.org/10.1103/PhysRevLett.123.123402} {\bibfield  {journal}
  {\bibinfo  {journal} {Phys. Rev. Lett.}\ }\textbf {\bibinfo {volume} {123}},\
  \bibinfo {pages} {123402} (\bibinfo {year} {2019})}\BibitemShut {NoStop}%
\bibitem [{\citenamefont {Micheli}\ \emph {et~al.}(2010)\citenamefont
  {Micheli}, \citenamefont {Idziaszek}, \citenamefont {Pupillo}, \citenamefont
  {Baranov}, \citenamefont {Zoller},\ and\ \citenamefont
  {Julienne}}]{PhysRevLett.105.073202}%
  \BibitemOpen
  \bibfield  {author} {\bibinfo {author} {\bibfnamefont {A.}~\bibnamefont
  {Micheli}}, \bibinfo {author} {\bibfnamefont {Z.}~\bibnamefont {Idziaszek}},
  \bibinfo {author} {\bibfnamefont {G.}~\bibnamefont {Pupillo}}, \bibinfo
  {author} {\bibfnamefont {M.~A.}\ \bibnamefont {Baranov}}, \bibinfo {author}
  {\bibfnamefont {P.}~\bibnamefont {Zoller}},\ and\ \bibinfo {author}
  {\bibfnamefont {P.~S.}\ \bibnamefont {Julienne}},\ }\bibfield  {title}
  {\bibinfo {title} {Universal rates for reactive ultracold polar molecules in
  reduced dimensions},\ }\href {https://doi.org/10.1103/PhysRevLett.105.073202}
  {\bibfield  {journal} {\bibinfo  {journal} {Phys. Rev. Lett.}\ }\textbf
  {\bibinfo {volume} {105}},\ \bibinfo {pages} {073202} (\bibinfo {year}
  {2010})}\BibitemShut {NoStop}%
\bibitem [{\citenamefont {Qu\'em\'ener}\ and\ \citenamefont
  {Bohn}(2011)}]{PhysRevA.83.012705}%
  \BibitemOpen
  \bibfield  {author} {\bibinfo {author} {\bibfnamefont {G.}~\bibnamefont
  {Qu\'em\'ener}}\ and\ \bibinfo {author} {\bibfnamefont {J.~L.}\ \bibnamefont
  {Bohn}},\ }\bibfield  {title} {\bibinfo {title} {Dynamics of ultracold
  molecules in confined geometry and electric field},\ }\href
  {https://doi.org/10.1103/PhysRevA.83.012705} {\bibfield  {journal} {\bibinfo
  {journal} {Phys. Rev. A}\ }\textbf {\bibinfo {volume} {83}},\ \bibinfo
  {pages} {012705} (\bibinfo {year} {2011})}\BibitemShut {NoStop}%
\bibitem [{\citenamefont {Zhu}\ \emph {et~al.}(2013)\citenamefont {Zhu},
  \citenamefont {Qu\'em\'ener}, \citenamefont {Rey},\ and\ \citenamefont
  {Holland}}]{PhysRevA.88.063405}%
  \BibitemOpen
  \bibfield  {author} {\bibinfo {author} {\bibfnamefont {B.}~\bibnamefont
  {Zhu}}, \bibinfo {author} {\bibfnamefont {G.}~\bibnamefont {Qu\'em\'ener}},
  \bibinfo {author} {\bibfnamefont {A.~M.}\ \bibnamefont {Rey}},\ and\ \bibinfo
  {author} {\bibfnamefont {M.~J.}\ \bibnamefont {Holland}},\ }\bibfield
  {title} {\bibinfo {title} {Evaporative cooling of reactive polar molecules
  confined in a two-dimensional geometry},\ }\href
  {https://doi.org/10.1103/PhysRevA.88.063405} {\bibfield  {journal} {\bibinfo
  {journal} {Phys. Rev. A}\ }\textbf {\bibinfo {volume} {88}},\ \bibinfo
  {pages} {063405} (\bibinfo {year} {2013})}\BibitemShut {NoStop}%
\bibitem [{\citenamefont {Valtolina}\ \emph {et~al.}(2020)\citenamefont
  {Valtolina}, \citenamefont {Matsuda}, \citenamefont {Tobias}, \citenamefont
  {Li}, \citenamefont {Marco},\ and\ \citenamefont {Ye}}]{2007.12277v1}%
  \BibitemOpen
  \bibfield  {author} {\bibinfo {author} {\bibfnamefont {G.}~\bibnamefont
  {Valtolina}}, \bibinfo {author} {\bibfnamefont {K.}~\bibnamefont {Matsuda}},
  \bibinfo {author} {\bibfnamefont {W.~G.}\ \bibnamefont {Tobias}}, \bibinfo
  {author} {\bibfnamefont {J.-R.}\ \bibnamefont {Li}}, \bibinfo {author}
  {\bibfnamefont {L.~D.}\ \bibnamefont {Marco}},\ and\ \bibinfo {author}
  {\bibfnamefont {J.}~\bibnamefont {Ye}},\ }\href@noop {} {\bibinfo {title}
  {Dipolar evaporation of reactive molecules to below the fermi temperature}}
  (\bibinfo {year} {2020}),\ \Eprint {https://arxiv.org/abs/2007.12277}
  {arXiv:2007.12277 [cond-mat.quant-gas]} \BibitemShut {NoStop}%
\bibitem [{\citenamefont {Kitagawa}\ and\ \citenamefont
  {Ueda}(1993)}]{Kitagawa1993}%
  \BibitemOpen
  \bibfield  {author} {\bibinfo {author} {\bibfnamefont {M.}~\bibnamefont
  {Kitagawa}}\ and\ \bibinfo {author} {\bibfnamefont {M.}~\bibnamefont
  {Ueda}},\ }\bibfield  {title} {\bibinfo {title} {Squeezed spin states},\
  }\href {https://doi.org/10.1103/PhysRevA.47.5138} {\bibfield  {journal}
  {\bibinfo  {journal} {Phys. Rev. A}\ }\textbf {\bibinfo {volume} {47}},\
  \bibinfo {pages} {5138} (\bibinfo {year} {1993})}\BibitemShut {NoStop}%
\bibitem [{\citenamefont {Pezz\'e}\ and\ \citenamefont
  {Smerzi}(2009)}]{Entanglement_OAT_2009}%
  \BibitemOpen
  \bibfield  {author} {\bibinfo {author} {\bibfnamefont {L.}~\bibnamefont
  {Pezz\'e}}\ and\ \bibinfo {author} {\bibfnamefont {A.}~\bibnamefont
  {Smerzi}},\ }\bibfield  {title} {\bibinfo {title} {Entanglement, nonlinear
  dynamics, and the heisenberg limit},\ }\href
  {https://doi.org/10.1103/PhysRevLett.102.100401} {\bibfield  {journal}
  {\bibinfo  {journal} {Phys. Rev. Lett.}\ }\textbf {\bibinfo {volume} {102}},\
  \bibinfo {pages} {100401} (\bibinfo {year} {2009})}\BibitemShut {NoStop}%
\bibitem [{\citenamefont {Sørensen}\ \emph {et~al.}(2001)\citenamefont
  {Sørensen}, \citenamefont {Duan}, \citenamefont {Cirac},\ and\ \citenamefont
  {Zoller}}]{Sorensen_OAT_BEC}%
  \BibitemOpen
  \bibfield  {author} {\bibinfo {author} {\bibfnamefont {A.}~\bibnamefont
  {Sørensen}}, \bibinfo {author} {\bibfnamefont {L.-M.}\ \bibnamefont {Duan}},
  \bibinfo {author} {\bibfnamefont {J.~I.}\ \bibnamefont {Cirac}},\ and\
  \bibinfo {author} {\bibfnamefont {P.}~\bibnamefont {Zoller}},\ }\bibfield
  {title} {\bibinfo {title} {Many-particle entanglement with bose–einstein
  condensates},\ }\href {https://doi.org/10.1038/35051038} {\bibfield
  {journal} {\bibinfo  {journal} {Nature}\ }\textbf {\bibinfo {volume} {409}},\
  \bibinfo {pages} {63–66} (\bibinfo {year} {2001})}\BibitemShut {NoStop}%
\bibitem [{\citenamefont {Schleier-Smith}\ \emph {et~al.}(2010)\citenamefont
  {Schleier-Smith}, \citenamefont {Leroux},\ and\ \citenamefont
  {Vuleti\ifmmode~\acute{c}\else \'{c}\fi{}}}]{Vladan_2010a}%
  \BibitemOpen
  \bibfield  {author} {\bibinfo {author} {\bibfnamefont {M.~H.}\ \bibnamefont
  {Schleier-Smith}}, \bibinfo {author} {\bibfnamefont {I.~D.}\ \bibnamefont
  {Leroux}},\ and\ \bibinfo {author} {\bibfnamefont {V.}~\bibnamefont
  {Vuleti\ifmmode~\acute{c}\else \'{c}\fi{}}},\ }\bibfield  {title} {\bibinfo
  {title} {States of an ensemble of two-level atoms with reduced quantum
  uncertainty},\ }\href {https://doi.org/10.1103/PhysRevLett.104.073604}
  {\bibfield  {journal} {\bibinfo  {journal} {Phys. Rev. Lett.}\ }\textbf
  {\bibinfo {volume} {104}},\ \bibinfo {pages} {073604} (\bibinfo {year}
  {2010})}\BibitemShut {NoStop}%
\bibitem [{\citenamefont {Leroux}\ \emph {et~al.}(2010)\citenamefont {Leroux},
  \citenamefont {Schleier-Smith},\ and\ \citenamefont
  {Vuleti\ifmmode~\acute{c}\else \'{c}\fi{}}}]{Vladan_2010b}%
  \BibitemOpen
  \bibfield  {author} {\bibinfo {author} {\bibfnamefont {I.~D.}\ \bibnamefont
  {Leroux}}, \bibinfo {author} {\bibfnamefont {M.~H.}\ \bibnamefont
  {Schleier-Smith}},\ and\ \bibinfo {author} {\bibfnamefont {V.}~\bibnamefont
  {Vuleti\ifmmode~\acute{c}\else \'{c}\fi{}}},\ }\bibfield  {title} {\bibinfo
  {title} {Orientation-dependent entanglement lifetime in a squeezed atomic
  clock},\ }\href {https://doi.org/10.1103/PhysRevLett.104.250801} {\bibfield
  {journal} {\bibinfo  {journal} {Phys. Rev. Lett.}\ }\textbf {\bibinfo
  {volume} {104}},\ \bibinfo {pages} {250801} (\bibinfo {year}
  {2010})}\BibitemShut {NoStop}%
\bibitem [{\citenamefont {Bohnet}\ \emph {et~al.}(2016)\citenamefont {Bohnet},
  \citenamefont {Sawyer}, \citenamefont {Britton}, \citenamefont {Wall},
  \citenamefont {Rey}, \citenamefont {Foss-Feig},\ and\ \citenamefont
  {Bollinger}}]{Ions_Bollinger_2016}%
  \BibitemOpen
  \bibfield  {author} {\bibinfo {author} {\bibfnamefont {J.~G.}\ \bibnamefont
  {Bohnet}}, \bibinfo {author} {\bibfnamefont {B.~C.}\ \bibnamefont {Sawyer}},
  \bibinfo {author} {\bibfnamefont {J.~W.}\ \bibnamefont {Britton}}, \bibinfo
  {author} {\bibfnamefont {M.~L.}\ \bibnamefont {Wall}}, \bibinfo {author}
  {\bibfnamefont {A.~M.}\ \bibnamefont {Rey}}, \bibinfo {author} {\bibfnamefont
  {M.}~\bibnamefont {Foss-Feig}},\ and\ \bibinfo {author} {\bibfnamefont
  {J.~J.}\ \bibnamefont {Bollinger}},\ }\bibfield  {title} {\bibinfo {title}
  {Quantum spin dynamics and entanglement generation with hundreds of trapped
  ions},\ }\href {https://doi.org/10.1126/science.aad9958} {\bibfield
  {journal} {\bibinfo  {journal} {Science}\ }\textbf {\bibinfo {volume}
  {352}},\ \bibinfo {pages} {1297–1301} (\bibinfo {year} {2016})}\BibitemShut
  {NoStop}%
\bibitem [{\citenamefont {Pedrozo-Peñafiel}\ \emph {et~al.}(2020)\citenamefont
  {Pedrozo-Peñafiel}, \citenamefont {Colombo}, \citenamefont {Shu},
  \citenamefont {Adiyatullin}, \citenamefont {Li}, \citenamefont {Mendez},
  \citenamefont {Braverman}, \citenamefont {Kawasaki}, \citenamefont
  {Akamatsu}, \citenamefont {Xiao},\ and\ \citenamefont
  {Vuletić}}]{Vladan_Clock_2020}%
  \BibitemOpen
  \bibfield  {author} {\bibinfo {author} {\bibfnamefont {E.}~\bibnamefont
  {Pedrozo-Peñafiel}}, \bibinfo {author} {\bibfnamefont {S.}~\bibnamefont
  {Colombo}}, \bibinfo {author} {\bibfnamefont {C.}~\bibnamefont {Shu}},
  \bibinfo {author} {\bibfnamefont {A.~F.}\ \bibnamefont {Adiyatullin}},
  \bibinfo {author} {\bibfnamefont {Z.}~\bibnamefont {Li}}, \bibinfo {author}
  {\bibfnamefont {E.}~\bibnamefont {Mendez}}, \bibinfo {author} {\bibfnamefont
  {B.}~\bibnamefont {Braverman}}, \bibinfo {author} {\bibfnamefont
  {A.}~\bibnamefont {Kawasaki}}, \bibinfo {author} {\bibfnamefont
  {D.}~\bibnamefont {Akamatsu}}, \bibinfo {author} {\bibfnamefont
  {Y.}~\bibnamefont {Xiao}},\ and\ \bibinfo {author} {\bibfnamefont
  {V.}~\bibnamefont {Vuletić}},\ }\href@noop {} {\bibinfo {title}
  {Entanglement-enhanced optical atomic clock}} (\bibinfo {year} {2020}),\
  \Eprint {https://arxiv.org/abs/2006.07501} {arXiv:2006.07501 [quant-ph]}
  \BibitemShut {NoStop}%
\bibitem [{\citenamefont {Davis}\ \emph {et~al.}(2016)\citenamefont {Davis},
  \citenamefont {Bentsen},\ and\ \citenamefont
  {Schleier-Smith}}]{Schleier_Smith_Heisenberg}%
  \BibitemOpen
  \bibfield  {author} {\bibinfo {author} {\bibfnamefont {E.}~\bibnamefont
  {Davis}}, \bibinfo {author} {\bibfnamefont {G.}~\bibnamefont {Bentsen}},\
  and\ \bibinfo {author} {\bibfnamefont {M.}~\bibnamefont {Schleier-Smith}},\
  }\bibfield  {title} {\bibinfo {title} {Approaching the heisenberg limit
  without single-particle detection},\ }\href
  {https://doi.org/10.1103/PhysRevLett.116.053601} {\bibfield  {journal}
  {\bibinfo  {journal} {Phys. Rev. Lett.}\ }\textbf {\bibinfo {volume} {116}},\
  \bibinfo {pages} {053601} (\bibinfo {year} {2016})}\BibitemShut {NoStop}%
\bibitem [{\citenamefont {Schulte}\ \emph {et~al.}(2020)\citenamefont
  {Schulte}, \citenamefont {Martínez-Lahuerta}, \citenamefont {Scharnagl},\
  and\ \citenamefont {Hammerer}}]{Schulte_2020}%
  \BibitemOpen
  \bibfield  {author} {\bibinfo {author} {\bibfnamefont {M.}~\bibnamefont
  {Schulte}}, \bibinfo {author} {\bibfnamefont {V.~J.}\ \bibnamefont
  {Martínez-Lahuerta}}, \bibinfo {author} {\bibfnamefont {M.~S.}\ \bibnamefont
  {Scharnagl}},\ and\ \bibinfo {author} {\bibfnamefont {K.}~\bibnamefont
  {Hammerer}},\ }\bibfield  {title} {\bibinfo {title} {Ramsey interferometry
  with generalized one-axis twisting echoes},\ }\href
  {https://doi.org/10.22331/q-2020-05-15-268} {\bibfield  {journal} {\bibinfo
  {journal} {Quantum}\ }\textbf {\bibinfo {volume} {4}},\ \bibinfo {pages}
  {268} (\bibinfo {year} {2020})}\BibitemShut {NoStop}%
\bibitem [{sup()}]{supplemental}%
  \BibitemOpen
  \href@noop {} {}\bibinfo {note} {See Supplemental Material at [URL will be
  inserted by publisher] for additional details on the derivation of the
  spin-model in the quasi-2D limit, the matrix elements of the interactions,
  the kinetic theory, and the full equations of motion including losses, which
  includes Refs. 5, 27, 41, 43, 51-56}\BibitemShut {NoStop}%
\bibitem [{\citenamefont {Deutsch}\ \emph {et~al.}(2010)\citenamefont
  {Deutsch}, \citenamefont {Ramirez-Martinez}, \citenamefont {Lacro\^ute},
  \citenamefont {Reinhard}, \citenamefont {Schneider}, \citenamefont {Fuchs},
  \citenamefont {Pi\'echon}, \citenamefont {Lalo\"e}, \citenamefont {Reichel},\
  and\ \citenamefont {Rosenbusch}}]{Deutsch_2010}%
  \BibitemOpen
  \bibfield  {author} {\bibinfo {author} {\bibfnamefont {C.}~\bibnamefont
  {Deutsch}}, \bibinfo {author} {\bibfnamefont {F.}~\bibnamefont
  {Ramirez-Martinez}}, \bibinfo {author} {\bibfnamefont {C.}~\bibnamefont
  {Lacro\^ute}}, \bibinfo {author} {\bibfnamefont {F.}~\bibnamefont
  {Reinhard}}, \bibinfo {author} {\bibfnamefont {T.}~\bibnamefont {Schneider}},
  \bibinfo {author} {\bibfnamefont {J.~N.}\ \bibnamefont {Fuchs}}, \bibinfo
  {author} {\bibfnamefont {F.}~\bibnamefont {Pi\'echon}}, \bibinfo {author}
  {\bibfnamefont {F.}~\bibnamefont {Lalo\"e}}, \bibinfo {author} {\bibfnamefont
  {J.}~\bibnamefont {Reichel}},\ and\ \bibinfo {author} {\bibfnamefont
  {P.}~\bibnamefont {Rosenbusch}},\ }\bibfield  {title} {\bibinfo {title} {Spin
  self-rephasing and very long coherence times in a trapped atomic ensemble},\
  }\href {https://doi.org/10.1103/PhysRevLett.105.020401} {\bibfield  {journal}
  {\bibinfo  {journal} {Phys. Rev. Lett.}\ }\textbf {\bibinfo {volume} {105}},\
  \bibinfo {pages} {020401} (\bibinfo {year} {2010})}\BibitemShut {NoStop}%
\bibitem [{\citenamefont {Smale}\ \emph {et~al.}(2019)\citenamefont {Smale},
  \citenamefont {He}, \citenamefont {Olsen}, \citenamefont {Jackson},
  \citenamefont {Sharum}, \citenamefont {Trotzky}, \citenamefont {Marino},
  \citenamefont {Rey},\ and\ \citenamefont {Thywissen}}]{Smaleeaax2019}%
  \BibitemOpen
  \bibfield  {author} {\bibinfo {author} {\bibfnamefont {S.}~\bibnamefont
  {Smale}}, \bibinfo {author} {\bibfnamefont {P.}~\bibnamefont {He}}, \bibinfo
  {author} {\bibfnamefont {B.~A.}\ \bibnamefont {Olsen}}, \bibinfo {author}
  {\bibfnamefont {K.~G.}\ \bibnamefont {Jackson}}, \bibinfo {author}
  {\bibfnamefont {H.}~\bibnamefont {Sharum}}, \bibinfo {author} {\bibfnamefont
  {S.}~\bibnamefont {Trotzky}}, \bibinfo {author} {\bibfnamefont
  {J.}~\bibnamefont {Marino}}, \bibinfo {author} {\bibfnamefont {A.~M.}\
  \bibnamefont {Rey}},\ and\ \bibinfo {author} {\bibfnamefont {J.~H.}\
  \bibnamefont {Thywissen}},\ }\bibfield  {title} {\bibinfo {title}
  {Observation of a transition between dynamical phases in a quantum degenerate
  fermi gas},\ }\href {https://doi.org/10.1126/sciadv.aax1568} {\bibfield
  {journal} {\bibinfo  {journal} {Science Advances}\ }\textbf {\bibinfo
  {volume} {5}} (\bibinfo {year} {2019})}\BibitemShut {NoStop}%
\bibitem [{\citenamefont {Rey}\ \emph {et~al.}(2014)\citenamefont {Rey},
  \citenamefont {Gorshkov}, \citenamefont {Kraus}, \citenamefont {Martin},
  \citenamefont {Bishof}, \citenamefont {Swallows}, \citenamefont {Zhang},
  \citenamefont {Benko}, \citenamefont {Ye}, \citenamefont {Lemke},\ and\
  \citenamefont {Ludlow}}]{martin2013}%
  \BibitemOpen
  \bibfield  {author} {\bibinfo {author} {\bibfnamefont {A.~M.}\ \bibnamefont
  {Rey}}, \bibinfo {author} {\bibfnamefont {A.~V.}\ \bibnamefont {Gorshkov}},
  \bibinfo {author} {\bibfnamefont {C.~V.}\ \bibnamefont {Kraus}}, \bibinfo
  {author} {\bibfnamefont {M.~J.}\ \bibnamefont {Martin}}, \bibinfo {author}
  {\bibfnamefont {M.}~\bibnamefont {Bishof}}, \bibinfo {author} {\bibfnamefont
  {M.~D.}\ \bibnamefont {Swallows}}, \bibinfo {author} {\bibfnamefont
  {X.}~\bibnamefont {Zhang}}, \bibinfo {author} {\bibfnamefont
  {C.}~\bibnamefont {Benko}}, \bibinfo {author} {\bibfnamefont
  {J.}~\bibnamefont {Ye}}, \bibinfo {author} {\bibfnamefont {N.~D.}\
  \bibnamefont {Lemke}},\ and\ \bibinfo {author} {\bibfnamefont {A.~D.}\
  \bibnamefont {Ludlow}},\ }\bibfield  {title} {\bibinfo {title} {Probing
  many-body interactions in an optical lattice clock},\ }\href
  {https://doi.org/https://doi.org/10.1016/j.aop.2013.11.002} {\bibfield
  {journal} {\bibinfo  {journal} {Annals of Physics}\ }\textbf {\bibinfo
  {volume} {340}},\ \bibinfo {pages} {311} (\bibinfo {year}
  {2014})}\BibitemShut {NoStop}%
\bibitem [{\citenamefont {Gorshkov}\ \emph {et~al.}(2008)\citenamefont
  {Gorshkov}, \citenamefont {Rabl}, \citenamefont {Pupillo}, \citenamefont
  {Micheli}, \citenamefont {Zoller}, \citenamefont {Lukin},\ and\ \citenamefont
  {B\"uchler}}]{Buechler2008}%
  \BibitemOpen
  \bibfield  {author} {\bibinfo {author} {\bibfnamefont {A.~V.}\ \bibnamefont
  {Gorshkov}}, \bibinfo {author} {\bibfnamefont {P.}~\bibnamefont {Rabl}},
  \bibinfo {author} {\bibfnamefont {G.}~\bibnamefont {Pupillo}}, \bibinfo
  {author} {\bibfnamefont {A.}~\bibnamefont {Micheli}}, \bibinfo {author}
  {\bibfnamefont {P.}~\bibnamefont {Zoller}}, \bibinfo {author} {\bibfnamefont
  {M.~D.}\ \bibnamefont {Lukin}},\ and\ \bibinfo {author} {\bibfnamefont
  {H.~P.}\ \bibnamefont {B\"uchler}},\ }\bibfield  {title} {\bibinfo {title}
  {Suppression of inelastic collisions between polar molecules with a repulsive
  shield},\ }\href {https://doi.org/10.1103/PhysRevLett.101.073201} {\bibfield
  {journal} {\bibinfo  {journal} {Phys. Rev. Lett.}\ }\textbf {\bibinfo
  {volume} {101}},\ \bibinfo {pages} {073201} (\bibinfo {year}
  {2008})}\BibitemShut {NoStop}%
\bibitem [{\citenamefont {Neyenhuis}\ \emph {et~al.}(2012)\citenamefont
  {Neyenhuis}, \citenamefont {Yan}, \citenamefont {Moses}, \citenamefont
  {Covey}, \citenamefont {Chotia}, \citenamefont {Petrov}, \citenamefont
  {Kotochigova}, \citenamefont {Ye},\ and\ \citenamefont
  {Jin}}]{Neyenhuis2012}%
  \BibitemOpen
  \bibfield  {author} {\bibinfo {author} {\bibfnamefont {B.}~\bibnamefont
  {Neyenhuis}}, \bibinfo {author} {\bibfnamefont {B.}~\bibnamefont {Yan}},
  \bibinfo {author} {\bibfnamefont {S.~A.}\ \bibnamefont {Moses}}, \bibinfo
  {author} {\bibfnamefont {J.~P.}\ \bibnamefont {Covey}}, \bibinfo {author}
  {\bibfnamefont {A.}~\bibnamefont {Chotia}}, \bibinfo {author} {\bibfnamefont
  {A.}~\bibnamefont {Petrov}}, \bibinfo {author} {\bibfnamefont
  {S.}~\bibnamefont {Kotochigova}}, \bibinfo {author} {\bibfnamefont
  {J.}~\bibnamefont {Ye}},\ and\ \bibinfo {author} {\bibfnamefont {D.~S.}\
  \bibnamefont {Jin}},\ }\bibfield  {title} {\bibinfo {title} {Anisotropic
  polarizability of ultracold polar $^{40}\mathrm{K}^{87}\mathrm{Rb}$
  molecules},\ }\href {https://doi.org/10.1103/PhysRevLett.109.230403}
  {\bibfield  {journal} {\bibinfo  {journal} {Phys. Rev. Lett.}\ }\textbf
  {\bibinfo {volume} {109}},\ \bibinfo {pages} {230403} (\bibinfo {year}
  {2012})}\BibitemShut {NoStop}%
\bibitem [{\citenamefont {Schachenmayer}\ \emph {et~al.}(2015)\citenamefont
  {Schachenmayer}, \citenamefont {Pikovski},\ and\ \citenamefont
  {Rey}}]{Schachenmayer2015}%
  \BibitemOpen
  \bibfield  {author} {\bibinfo {author} {\bibfnamefont {J.}~\bibnamefont
  {Schachenmayer}}, \bibinfo {author} {\bibfnamefont {A.}~\bibnamefont
  {Pikovski}},\ and\ \bibinfo {author} {\bibfnamefont {A.~M.}\ \bibnamefont
  {Rey}},\ }\bibfield  {title} {\bibinfo {title} {Many-body quantum spin
  dynamics with monte carlo trajectories on a discrete phase space},\ }\href
  {https://doi.org/10.1103/PhysRevX.5.011022} {\bibfield  {journal} {\bibinfo
  {journal} {Phys. Rev. X}\ }\textbf {\bibinfo {volume} {5}},\ \bibinfo {pages}
  {011022} (\bibinfo {year} {2015})}\BibitemShut {NoStop}%
\bibitem [{\citenamefont {Zhu}\ \emph {et~al.}(2019)\citenamefont {Zhu},
  \citenamefont {Rey},\ and\ \citenamefont {Schachenmayer}}]{Zhu_2019}%
  \BibitemOpen
  \bibfield  {author} {\bibinfo {author} {\bibfnamefont {B.}~\bibnamefont
  {Zhu}}, \bibinfo {author} {\bibfnamefont {A.~M.}\ \bibnamefont {Rey}},\ and\
  \bibinfo {author} {\bibfnamefont {J.}~\bibnamefont {Schachenmayer}},\
  }\bibfield  {title} {\bibinfo {title} {A generalized phase space approach for
  solving quantum spin dynamics},\ }\href
  {https://doi.org/10.1088/1367-2630/ab354d} {\bibfield  {journal} {\bibinfo
  {journal} {New Journal of Physics}\ }\textbf {\bibinfo {volume} {21}},\
  \bibinfo {pages} {082001} (\bibinfo {year} {2019})}\BibitemShut {NoStop}%
\bibitem [{\citenamefont {Rey}\ \emph {et~al.}(2008)\citenamefont {Rey},
  \citenamefont {Jiang}, \citenamefont {Fleischhauer}, \citenamefont {Demler},\
  and\ \citenamefont {Lukin}}]{Rey2008}%
  \BibitemOpen
  \bibfield  {author} {\bibinfo {author} {\bibfnamefont {A.~M.}\ \bibnamefont
  {Rey}}, \bibinfo {author} {\bibfnamefont {L.}~\bibnamefont {Jiang}}, \bibinfo
  {author} {\bibfnamefont {M.}~\bibnamefont {Fleischhauer}}, \bibinfo {author}
  {\bibfnamefont {E.}~\bibnamefont {Demler}},\ and\ \bibinfo {author}
  {\bibfnamefont {M.~D.}\ \bibnamefont {Lukin}},\ }\bibfield  {title} {\bibinfo
  {title} {Many-body protected entanglement generation in interacting spin
  systems},\ }\href {https://doi.org/10.1103/PhysRevA.77.052305} {\bibfield
  {journal} {\bibinfo  {journal} {Phys. Rev. A}\ }\textbf {\bibinfo {volume}
  {77}},\ \bibinfo {pages} {052305} (\bibinfo {year} {2008})}\BibitemShut
  {NoStop}%
\bibitem [{\citenamefont {Wineland}\ \emph {et~al.}(1992)\citenamefont
  {Wineland}, \citenamefont {Bollinger}, \citenamefont {Itano}, \citenamefont
  {Moore},\ and\ \citenamefont {Heinzen}}]{Wineland1992}%
  \BibitemOpen
  \bibfield  {author} {\bibinfo {author} {\bibfnamefont {D.~J.}\ \bibnamefont
  {Wineland}}, \bibinfo {author} {\bibfnamefont {J.~J.}\ \bibnamefont
  {Bollinger}}, \bibinfo {author} {\bibfnamefont {W.~M.}\ \bibnamefont
  {Itano}}, \bibinfo {author} {\bibfnamefont {F.~L.}\ \bibnamefont {Moore}},\
  and\ \bibinfo {author} {\bibfnamefont {D.~J.}\ \bibnamefont {Heinzen}},\
  }\bibfield  {title} {\bibinfo {title} {Spin squeezing and reduced quantum
  noise in spectroscopy},\ }\href {https://doi.org/10.1103/PhysRevA.46.R6797}
  {\bibfield  {journal} {\bibinfo  {journal} {Phys. Rev. A}\ }\textbf {\bibinfo
  {volume} {46}},\ \bibinfo {pages} {R6797} (\bibinfo {year}
  {1992})}\BibitemShut {NoStop}%
\bibitem [{\citenamefont {Wineland}\ \emph {et~al.}(1994)\citenamefont
  {Wineland}, \citenamefont {Bollinger}, \citenamefont {Itano},\ and\
  \citenamefont {Heinzen}}]{Wineland1994}%
  \BibitemOpen
  \bibfield  {author} {\bibinfo {author} {\bibfnamefont {D.~J.}\ \bibnamefont
  {Wineland}}, \bibinfo {author} {\bibfnamefont {J.~J.}\ \bibnamefont
  {Bollinger}}, \bibinfo {author} {\bibfnamefont {W.~M.}\ \bibnamefont
  {Itano}},\ and\ \bibinfo {author} {\bibfnamefont {D.~J.}\ \bibnamefont
  {Heinzen}},\ }\bibfield  {title} {\bibinfo {title} {Squeezed atomic states
  and projection noise in spectroscopy},\ }\href
  {https://doi.org/10.1103/PhysRevA.50.67} {\bibfield  {journal} {\bibinfo
  {journal} {Phys. Rev. A}\ }\textbf {\bibinfo {volume} {50}},\ \bibinfo
  {pages} {67} (\bibinfo {year} {1994})}\BibitemShut {NoStop}%
\bibitem [{\citenamefont {He}\ \emph {et~al.}(2019)\citenamefont {He},
  \citenamefont {Perlin}, \citenamefont {Muleady}, \citenamefont {Lewis-Swan},
  \citenamefont {Hutson}, \citenamefont {Ye},\ and\ \citenamefont
  {Rey}}]{PhysRevResearch.1.033075}%
  \BibitemOpen
  \bibfield  {author} {\bibinfo {author} {\bibfnamefont {P.}~\bibnamefont
  {He}}, \bibinfo {author} {\bibfnamefont {M.~A.}\ \bibnamefont {Perlin}},
  \bibinfo {author} {\bibfnamefont {S.~R.}\ \bibnamefont {Muleady}}, \bibinfo
  {author} {\bibfnamefont {R.~J.}\ \bibnamefont {Lewis-Swan}}, \bibinfo
  {author} {\bibfnamefont {R.~B.}\ \bibnamefont {Hutson}}, \bibinfo {author}
  {\bibfnamefont {J.}~\bibnamefont {Ye}},\ and\ \bibinfo {author}
  {\bibfnamefont {A.~M.}\ \bibnamefont {Rey}},\ }\bibfield  {title} {\bibinfo
  {title} {Engineering spin squeezing in a 3d optical lattice with interacting
  spin-orbit-coupled fermions},\ }\href
  {https://doi.org/10.1103/PhysRevResearch.1.033075} {\bibfield  {journal}
  {\bibinfo  {journal} {Phys. Rev. Research}\ }\textbf {\bibinfo {volume}
  {1}},\ \bibinfo {pages} {033075} (\bibinfo {year} {2019})}\BibitemShut
  {NoStop}%
\bibitem [{\citenamefont {See\ss{}elberg}\ \emph {et~al.}(2018)\citenamefont
  {See\ss{}elberg}, \citenamefont {Luo}, \citenamefont {Li}, \citenamefont
  {Bause}, \citenamefont {Kotochigova}, \citenamefont {Bloch},\ and\
  \citenamefont {Gohle}}]{PhysRevLett.121.253401}%
  \BibitemOpen
  \bibfield  {author} {\bibinfo {author} {\bibfnamefont {F.}~\bibnamefont
  {See\ss{}elberg}}, \bibinfo {author} {\bibfnamefont {X.-Y.}\ \bibnamefont
  {Luo}}, \bibinfo {author} {\bibfnamefont {M.}~\bibnamefont {Li}}, \bibinfo
  {author} {\bibfnamefont {R.}~\bibnamefont {Bause}}, \bibinfo {author}
  {\bibfnamefont {S.}~\bibnamefont {Kotochigova}}, \bibinfo {author}
  {\bibfnamefont {I.}~\bibnamefont {Bloch}},\ and\ \bibinfo {author}
  {\bibfnamefont {C.}~\bibnamefont {Gohle}},\ }\bibfield  {title} {\bibinfo
  {title} {Extending rotational coherence of interacting polar molecules in a
  spin-decoupled magic trap},\ }\href
  {https://doi.org/10.1103/PhysRevLett.121.253401} {\bibfield  {journal}
  {\bibinfo  {journal} {Phys. Rev. Lett.}\ }\textbf {\bibinfo {volume} {121}},\
  \bibinfo {pages} {253401} (\bibinfo {year} {2018})}\BibitemShut {NoStop}%
\bibitem [{\citenamefont {Affolter}\ \emph {et~al.}(2020)\citenamefont
  {Affolter}, \citenamefont {Gilmore}, \citenamefont {Jordan},\ and\
  \citenamefont {Bollinger}}]{Bollinger_2020}%
  \BibitemOpen
  \bibfield  {author} {\bibinfo {author} {\bibfnamefont {M.}~\bibnamefont
  {Affolter}}, \bibinfo {author} {\bibfnamefont {K.~A.}\ \bibnamefont
  {Gilmore}}, \bibinfo {author} {\bibfnamefont {J.~E.}\ \bibnamefont
  {Jordan}},\ and\ \bibinfo {author} {\bibfnamefont {J.~J.}\ \bibnamefont
  {Bollinger}},\ }\bibfield  {title} {\bibinfo {title} {Phase-coherent sensing
  of the center-of-mass motion of trapped-ion crystals},\ }\href
  {https://doi.org/10.1103/PhysRevA.102.052609} {\bibfield  {journal} {\bibinfo
   {journal} {Phys. Rev. A}\ }\textbf {\bibinfo {volume} {102}},\ \bibinfo
  {pages} {052609} (\bibinfo {year} {2020})}\BibitemShut {NoStop}%
\bibitem [{\citenamefont {Jing}\ \emph {et~al.}(2020)\citenamefont {Jing},
  \citenamefont {Hu}, \citenamefont {Ma}, \citenamefont {Zhang}, \citenamefont
  {Zhang}, \citenamefont {Xiao},\ and\ \citenamefont {Jia}}]{Jing2020}%
  \BibitemOpen
  \bibfield  {author} {\bibinfo {author} {\bibfnamefont {M.}~\bibnamefont
  {Jing}}, \bibinfo {author} {\bibfnamefont {Y.}~\bibnamefont {Hu}}, \bibinfo
  {author} {\bibfnamefont {J.}~\bibnamefont {Ma}}, \bibinfo {author}
  {\bibfnamefont {H.}~\bibnamefont {Zhang}}, \bibinfo {author} {\bibfnamefont
  {L.}~\bibnamefont {Zhang}}, \bibinfo {author} {\bibfnamefont
  {L.}~\bibnamefont {Xiao}},\ and\ \bibinfo {author} {\bibfnamefont
  {S.}~\bibnamefont {Jia}},\ }\bibfield  {title} {\bibinfo {title} {Atomic
  superheterodyne receiver based on microwave-dressed rydberg spectroscopy},\
  }\href {https://doi.org/10.1038/s41567-020-0918-5} {\bibfield  {journal}
  {\bibinfo  {journal} {Nature Physics}\ }\textbf {\bibinfo {volume} {16}},\
  \bibinfo {pages} {911–915} (\bibinfo {year} {2020})}\BibitemShut {NoStop}%
\bibitem [{\citenamefont {Truman}\ and\ \citenamefont
  {Zhao}(1999)}]{Truman1999}%
  \BibitemOpen
  \bibfield  {author} {\bibinfo {author} {\bibfnamefont {A.}~\bibnamefont
  {Truman}}\ and\ \bibinfo {author} {\bibfnamefont {H.~Z.}\ \bibnamefont
  {Zhao}},\ }\bibfield  {title} {\bibinfo {title} {Semi-classical limit of wave
  functions},\ }\href {https://doi.org/10.1090/s0002-9939-99-05469-6}
  {\bibfield  {journal} {\bibinfo  {journal} {Proceedings of the American
  Mathematical Society}\ }\textbf {\bibinfo {volume} {128}},\ \bibinfo {pages}
  {1003} (\bibinfo {year} {1999})}\BibitemShut {NoStop}%
\bibitem [{\citenamefont {Bruun}(2012)}]{Bruun_2012}%
  \BibitemOpen
  \bibfield  {author} {\bibinfo {author} {\bibfnamefont {G.~M.}\ \bibnamefont
  {Bruun}},\ }\bibfield  {title} {\bibinfo {title} {Shear viscosity and
  spin-diffusion coefficient of a two-dimensional fermi gas},\ }\href
  {https://doi.org/10.1103/PhysRevA.85.013636} {\bibfield  {journal} {\bibinfo
  {journal} {Phys. Rev. A}\ }\textbf {\bibinfo {volume} {85}},\ \bibinfo
  {pages} {013636} (\bibinfo {year} {2012})}\BibitemShut {NoStop}%
\bibitem [{\citenamefont {Sch{\"a}fer}(2012)}]{Schafer_2012}%
  \BibitemOpen
  \bibfield  {author} {\bibinfo {author} {\bibfnamefont {T.}~\bibnamefont
  {Sch{\"a}fer}},\ }\bibfield  {title} {\bibinfo {title} {Shear viscosity and
  damping of collective modes in a two-dimensional fermi gas},\ }\href
  {https://doi.org/10.1103/PhysRevA.85.033623} {\bibfield  {journal} {\bibinfo
  {journal} {Phys. Rev. A}\ }\textbf {\bibinfo {volume} {85}},\ \bibinfo
  {pages} {033623} (\bibinfo {year} {2012})}\BibitemShut {NoStop}%
\bibitem [{\citenamefont {Enss}\ \emph {et~al.}(2012)\citenamefont {Enss},
  \citenamefont {K{\"u}ppersbusch},\ and\ \citenamefont {Fritz}}]{Enss_2012}%
  \BibitemOpen
  \bibfield  {author} {\bibinfo {author} {\bibfnamefont {T.}~\bibnamefont
  {Enss}}, \bibinfo {author} {\bibfnamefont {C.}~\bibnamefont
  {K{\"u}ppersbusch}},\ and\ \bibinfo {author} {\bibfnamefont {L.}~\bibnamefont
  {Fritz}},\ }\bibfield  {title} {\bibinfo {title} {Shear viscosity and spin
  diffusion in a two-dimensional fermi gas},\ }\href
  {https://doi.org/10.1103/PhysRevA.86.013617} {\bibfield  {journal} {\bibinfo
  {journal} {Phys. Rev. A}\ }\textbf {\bibinfo {volume} {86}},\ \bibinfo
  {pages} {013617} (\bibinfo {year} {2012})}\BibitemShut {NoStop}%
\bibitem [{\citenamefont {Baur}\ \emph {et~al.}(2013)\citenamefont {Baur},
  \citenamefont {Vogt}, \citenamefont {K{\"o}hl},\ and\ \citenamefont
  {Bruun}}]{Baur_2013}%
  \BibitemOpen
  \bibfield  {author} {\bibinfo {author} {\bibfnamefont {S.~K.}\ \bibnamefont
  {Baur}}, \bibinfo {author} {\bibfnamefont {E.}~\bibnamefont {Vogt}}, \bibinfo
  {author} {\bibfnamefont {M.}~\bibnamefont {K{\"o}hl}},\ and\ \bibinfo
  {author} {\bibfnamefont {G.~M.}\ \bibnamefont {Bruun}},\ }\bibfield  {title}
  {\bibinfo {title} {Collective modes of a two-dimensional spin-$1/2$ fermi gas
  in a harmonic trap},\ }\href {https://doi.org/10.1103/PhysRevA.87.043612}
  {\bibfield  {journal} {\bibinfo  {journal} {Phys. Rev. A}\ }\textbf {\bibinfo
  {volume} {87}},\ \bibinfo {pages} {043612} (\bibinfo {year}
  {2013})}\BibitemShut {NoStop}%
\bibitem [{\citenamefont {Bohn}\ \emph {et~al.}(2009)\citenamefont {Bohn},
  \citenamefont {Cavagnero},\ and\ \citenamefont {Ticknor}}]{Bohn_2009}%
  \BibitemOpen
  \bibfield  {author} {\bibinfo {author} {\bibfnamefont {J.~L.}\ \bibnamefont
  {Bohn}}, \bibinfo {author} {\bibfnamefont {M.}~\bibnamefont {Cavagnero}},\
  and\ \bibinfo {author} {\bibfnamefont {C.}~\bibnamefont {Ticknor}},\
  }\bibfield  {title} {\bibinfo {title} {Quasi-universal dipolar scattering in
  cold and ultracold gases},\ }\href
  {https://doi.org/10.1088/1367-2630/11/5/055039} {\bibfield  {journal}
  {\bibinfo  {journal} {New Journal of Physics}\ }\textbf {\bibinfo {volume}
  {11}},\ \bibinfo {pages} {055039} (\bibinfo {year} {2009})}\BibitemShut
  {NoStop}%
\end{thebibliography}

\end{document}


\title{Supplemental Material for: Dynamical generation of spin squeezing in ultra-cold  dipolar molecules}

\author{Thomas Bilitewski}
\affiliation{\JILA}
\affiliation{\CTQM}
\author{Luigi De Marco}
\affiliation{\JILA}
\author{Jun-Ru Li}
\affiliation{\JILA}
\author{Kyle Matsuda}
\affiliation{\JILA}
\author{William G. Tobias}
\affiliation{\JILA}
\author{Giacomo Valtolina}
\affiliation{\JILA}
\author{Jun Ye}
\affiliation{\JILA}
\author{Ana Maria Rey}
\affiliation{\JILA}
\affiliation{\CTQM}

\date{\today}
\maketitle
 \setcounter{equation}{0} \setcounter{figure}{0} \setcounter{table}{0}
 \renewcommand{\theequation}{S\arabic{equation}}
 \renewcommand{\thefigure}{S\arabic{figure}} \renewcommand{\bibnumfmt}[1]{[S#1]}
 \renewcommand{\thesection}{S\arabic{section}}
 \renewcommand{\citenumfont}[1]{S#1}
\section{Full Hamiltonian}
The Hamiltonian describing dipolar molecules in an optical lattice is given
by $\hat{H}=\hat{H}_{rot} + \hat{H}_{sp} + \hat{H}_{dd}$, where $\hat{H}_{rot}$ describes the internal
rotational state structure, $\hat{H}_{sp}$ the kinetic energy and optical lattice potentials, and $\hat{H}_{dd}$ the dipolar
interactions between the molecules.

Following Ref.~\cite{Gorshkov_2011} we describe the internal level structure of a single molecule via the
Hamiltonian
\begin{equation}
\hat{H}_{rot} =  B \vect{\hat{N}}^2 - \hat{d}_0 E
\end{equation}
where $B$ is the rotational constant, $\vect{\hat{N}}$ the angular momentum
operator of the rotation of the molecule, and $\hat{d}_0 = \vect{\hat{d}} \cdot \vect{e}_z$ is the
projection of the  dipole moment operator along the electric field $E$ which is
oriented along $z$. For KRb the permanent dipole moment is $d = 0.566$ Debye, and $B/h = 1.114$ GHz.

\begin{figure}
  \begin{minipage}[t]{0.49\columnwidth}
   \includegraphics[width=\columnwidth]{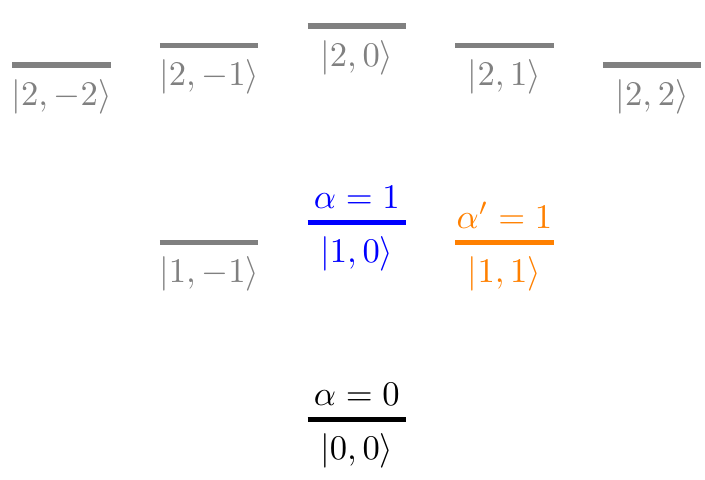}
  \end{minipage}
  \begin{minipage}[t]{0.49\columnwidth}
   \includegraphics[width=\columnwidth]{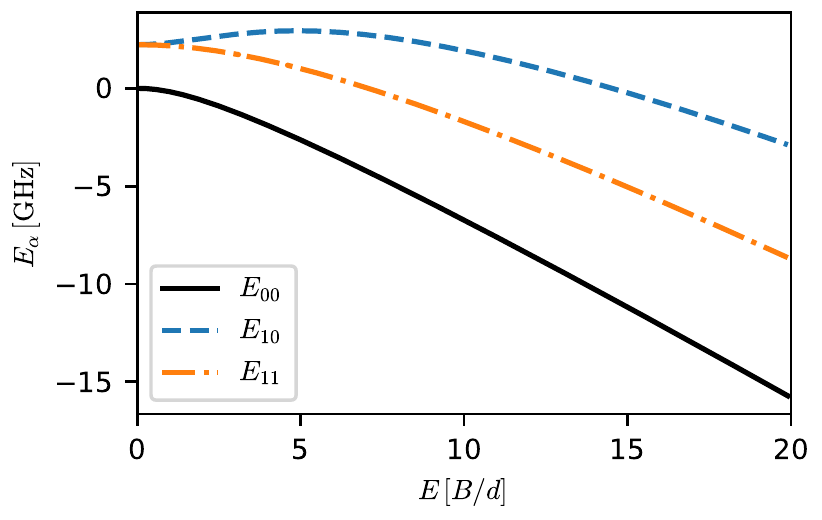}
  \end{minipage}
  \caption{a) Eigenstates of $H=B \vect{N}^2-d_0 E$, the states we consider are marked as $\alpha=0,1$ and $\alpha^{\prime}$.
          b) Eigenenergies of the $\ket{0,0}$, $\ket{1,0}$ and $\ket{1,1}$ states as a function of the applied external electric field $E$ in units of $B/d$ where for KRb $d = 0.566$ Debye, and $B/h = 1.114$ GHz, $B/d = 3.9$ kV/cm.
    \label{fig:state_structure}}
\end{figure}

We label the eigenstates of this Hamiltonian by the quantum numbers of the states they connect to in the limit of $E=0$ where $\vect{\hat{N}}^2 \ket{\mathcal{N},\mathcal{N}_Z} = \mathcal{N} (\mathcal{N}+1) \ket{\mathcal{N},\mathcal{N}_Z}$ and $\hat{N}_z \ket{\mathcal{N},\mathcal{N}_Z} = \mathcal{N}_Z \ket{\mathcal{N},\mathcal{N}_Z}$. Note that at non-zero $E$ states
with different $\mathcal{N}$ are mixed in the dressed eigenstates. We show a schematic structure
of the eigenstates and their eigenenergies in Fig.~\ref{fig:state_structure}.

\begin{figure}
  \begin{minipage}[t]{0.49\columnwidth}
    \includegraphics[width=.99\columnwidth]{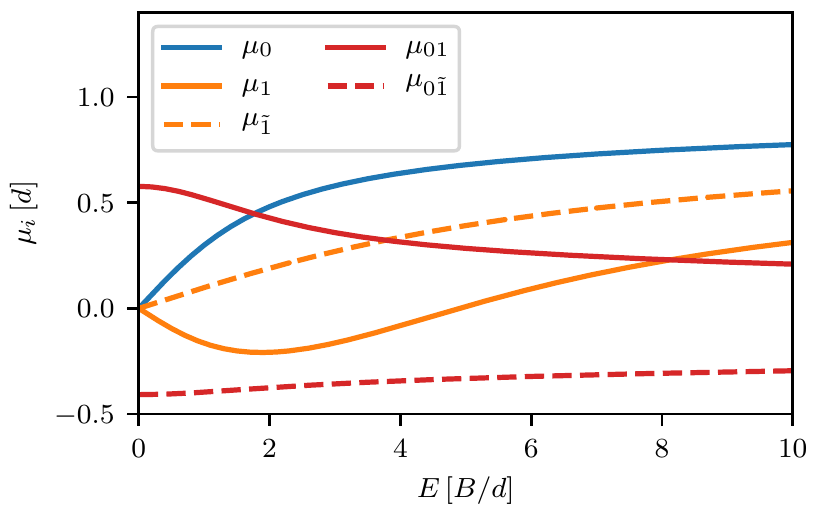}
  \end{minipage}
  \begin{minipage}[t]{0.49\columnwidth}
    \includegraphics[width=.99\columnwidth]{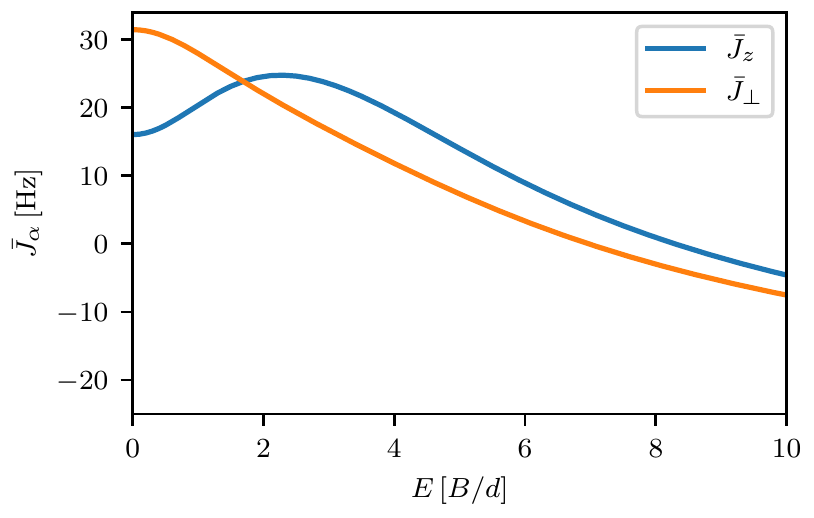}
  \end{minipage}
  \caption{a) Dipole moments $\mu_0 =\langle0,0|\hat{d}_0|0,0\rangle$, $\mu_1 =\langle1,0|\hat{d}_0|1,0\rangle$,
$\mu_{01} =\langle0,0|\hat{d}_0|1,0\rangle$ and $\mu_{\tilde{1}} = \expval{1,1}{\hat{d}_0}{1,1}$, $\mu_{0\tilde{1}} = \expval{0,0}{\hat{d}_-}{1,1}/\sqrt{2}$ in units of $d$ as a function of applied
electric field $E$ in units of $B/d$ evaluated for the basis described in the text.
b) Averaged couplings in the spin model $\bar{J}_z$ and $\bar{J}_{\perp}$, $\bar{J}_{\alpha}=1/N \sum_{i,j} J_{\alpha}(i,j)$ is the average over all occupied
states, as a function of the applied electric field for the $\ket{0,0}$ and $\ket{1,0}$ configuration.
For KRb $d = 0.566$ Debye, and $B/h = 1.114$ GHz, $B/d = 3.9$ kV/cm.
    \label{fig:dipole_moments}}
\end{figure}

For this work the relevant states are $ \ket{0,0}$ and $\ket{1,0}$ and $\ket{1,1}$.
For these states we show in Fig.~\ref{fig:dipole_moments}a) the non-vanishing dipole
moments $\mu_0 =\langle0,0|\hat{d}_0|0,0\rangle$, $\mu_1 =\langle1,0|\hat{d}_0|1,0\rangle$, $\mu_{01} =\langle0,0|\hat{d}_0|1,0\rangle$, and $\mu_{\tilde{1}} =\langle1,1|\hat{d}_0|1,1\rangle$, $\mu_{\tilde{1}0} = -\mu_{\tilde{1}0}=\langle1,1|\hat{d}_+|0,0\rangle/\sqrt{2}$, as a function of the applied electric field.

Since we always work in one of the spin 1/2 sub-spaces given by either $\ket{\downarrow}= \ket{0,0}$ and $\ket{\uparrow}=\ket{1,1}$ (basis I) or $\ket{\downarrow}= \ket{0,0}$ and $\ket{\uparrow}=\ket{1,0}$ (basis II) $\mu_{\uparrow}$, $\mu_{\downarrow}$ and $\mu_{\uparrow\downarrow}$ are understood to refer to the corresponding matrix elements just defined depending on the spin basis.

In this basis the Hamiltonian is written as 
  \begin{align}
    \hat{H}_{sp}  &= \int d^3\vect{r} \,\sum_{\alpha} \hat{\psi}^{\dagger}_{\alpha}(\vect{r}) \left(  -\frac{\nabla^2}{2m} +U^{3D}_{\alpha}(\vect{r}) \right) \hat{\psi}_{\alpha}(r) \\
  \hat{H}_{dd}  &= 1/2 \int \int d^3\vect{r} d^3 \vect{r}^{\prime} \, V^{3d}_{dd}(\vect{r}-\vect{r}^{\prime}) \times\Bigg\{\sum_{\alpha,\alpha^{\prime}} \mu_{\alpha} \mu_{\alpha^{\prime}} \hat{\psi}^{\dagger}_{\alpha}(\vect{r}) \hat{\psi}^{\dagger}_{\alpha^{\prime}}(\vect{r}^{\prime})  \hat{\psi}_{\alpha^{\prime}}(\vect{r}^{\prime}) \hat{\psi}_{\alpha}(\vect{r}) +\mu_{\downarrow\uparrow} \mu_{\uparrow\downarrow} \left[ \hat{\psi}^{\dagger}_{\downarrow}(\vect{r}) \hat{\psi}^{\dagger}_{\uparrow}(\vect{r}^{\prime})  \hat{\psi}_{\downarrow}(\vect{r}^{\prime}) \hat{\psi}_{\uparrow}(\vect{r})+\mathrm{H.c.} \right]  \Bigg\}
  \end{align}
  where $\hat{\psi}^{\dagger}_{\alpha}(\vect{r})$ creates a fermionic molecule at position $\vect{r}$ in internal state $\alpha=\uparrow,\downarrow$.

 The non-interacting Hamiltonian $H_{sp}$ describes the kinetic energy of the
 molecules subject to an internal state dependent trapping potential
 $U_{\alpha}(\vect{r})$, taken to be cylindrically symmetric and strongly
confining along the $Z$-direction, i.e. $U(\vect{r}) = 1/2 m \left(
  \omega^{\alpha} (X^2+Y^2) + \omega^{\alpha}_Z Z^2 \right)$ with
$\omega^{\alpha}_Z/\omega^{\alpha} \gg 1$.

The second line describes the elastic
interactions between molecular dipoles in states $\alpha$,$\alpha^{\prime}$
with the dipole elements defined above.
 We further assume that the dipoles are aligned along the $Z$-direction, e.g. by the applied electric field,
such that the dipole-dipole interactions depend on the spatial separation as
\begin{equation}
  V^{3d}_{dd}(\vect{r})=\frac{1}{4\pi r^5} (r^2 - 3Z^2)
 \end{equation}
  \subsection{Reduction to quasi-2D}
We will in the following consider the quasi-two-dimensional limit, where the
confinement along $Z$ sets the largest energy scale, e.g. $\hbar w_{\alpha,Z} \gg
\hbar w_{\alpha},\epsilon_F,k_{\rm B}T$. Then the fermionic molecules only occupy the lowest harmonic
oscillator state in the $Z$-direction, such that we can work with a reduced
model in 2D by integrating over the axial degrees of freedom.

In this limit we can express the field operator
as $\hat{\psi}(\vect{r}) = \hat{\psi}(X,Y) \phi_0(Z)$, where $\phi_0(Z) =
\frac{1}{(\sqrt{\pi} a_Z)^{1/2}}e^{-1/2 (Z/a_Z)^2}$ is the harmonic oscillator
ground state wavefunction along the $Z$ axis. Performing
the integrations over the $Z$ direction yields the two-dimensional Hamiltonian
\begin{equation}
  \begin{split}
  \hat{H}_{2D} &= \int d^2\vect{r} \,\sum_{\alpha} \hat{\psi}^{\dagger}_{\alpha}(\vect{r}) \left(  -\frac{\nabla^2}{2m} +U^{2D}_{\alpha}(\vect{r}) \right) \hat{\psi}_{\alpha}(r)  +1/2 \int \int d^2\vect{r} d^2 \vect{r}^{\prime} \, V^{2D}_{dd}(\vect{r}-\vect{r}^{\prime}) \\
  &\qquad \qquad \times\Bigg\{\sum_{\alpha,\alpha^{\prime}} \mu_{\alpha} \mu_{\alpha^{\prime}} \hat{\psi}^{\dagger}_{\alpha}(\vect{r}) \hat{\psi}^{\dagger}_{\alpha^{\prime}}(\vect{r}^{\prime})  \hat{\psi}_{\alpha^{\prime}}(\vect{r}^{\prime}) \hat{\psi}_{\alpha}(\vect{r})  +\mu_{\downarrow\uparrow}\mu_{\uparrow\downarrow} \left[ \hat{\psi}^{\dagger}_{\downarrow}(\vect{r}) \hat{\psi}^{\dagger}_{\uparrow}(\vect{r}^{\prime})  \hat{\psi}_{\downarrow}(\vect{r}^{\prime}) \hat{\psi}_{\uparrow}(\vect{r})+\mathrm{H.c.} \right]  \Bigg\}
  \end{split}
\end{equation}
where the trapping potential is just the in-plane part $U^{2D}_{\alpha}=1/2 m
\omega^{\alpha} (X^2+Y^2)$ and we dropped the zero-point energy of the harmonic oscillator along $Z$.

The reduced two-dimensional interaction is explicitly given by the following integral
\begin{equation}
V^{2D}_{dd}(\vect{r}-\vect{r}^{\prime})=\int \int dZ dZ^{\prime} V^{3D}_{dd}(\vect{r}-\vect{r}^{\prime},Z-Z^{\prime}) |\phi^0(Z)|^2|\phi^0(Z^{\prime})|^2 \, ,
\end{equation}
with the analytic result
\begin{equation}
    V^{2D}_{dd}(r) = \frac{1}{\sqrt{32 \pi^3}} \frac{1}{2 a_Z^3} e^{r^2/(4 a_Z^2)} \Bigg[ \left( 2+\frac{r^2}{a_Z^2} \right)  K_0\left(\frac{r^2}{4 a_Z^2} \right)   \quad -\frac{r^2}{a_Z^2} K_1 \left(\frac{r^2}{4 a_Z^2}\right) \Bigg]
 \end{equation}
 where $K_i$ is the modified Bessel function of the second kind.

 Correspondingly, we obtain in momentum space
 \begin{equation}
V^{2D}_{dd}(q) = \frac{1}{2a_Z} \left[\frac{2}{3} \sqrt{2/\pi} - q a_Z \mathrm{Erfc}\left( \frac{q a_Z}{\sqrt{2}} \right)  \right]
  \label{eq:Vdd_2D_q}
\end{equation}
where $\mathrm{Erfc}$ is the complementary errror function.

\subsection{Reduction to spin model}
In the next step we expand the field operator in a single-particle basis as
$\hat{\psi}(\vect{r}) = \sum_{i} \phi_i(\vect{r}) \hat{c}_i$, where $i$ enumerates the
single-particle states, and $\phi_i$ is the corresponding wavefunction, which we take to be the harmonic oscillator states of the weak transverse trap.

We further assume that the interactions
are sufficiently weak to not mix different single particle states $i \neq j$,
such that only the Hartree and Fock contributions of the interaction are
relevant. Finally, we assume that the single-particle states are either empty or
singly occupied, allowing us to use a pseudo-spin 1/2 representation defined by $ \hat{s}^{z}_{i}    = (\hat{c}_{\uparrow i}^{\dagger} \hat{c}_{\uparrow i}-\hat{c}_{0i}^{\dagger}\hat{c}_{0i})/2$, $\hat{s}_{i}^{+}  = \hat{c}_{\uparrow i}^{\dagger} \hat{c}_{\uparrow i}
$ and $ 1 = \hat{c}_{\uparrow i}^{\dagger}\hat{c}_{\uparrow i} + \hat{c}_{0i}^{\dagger}\hat{c}_{0i}$.

The single-particle part of the Hamiltonian then reduces to
\begin{equation}
\hat{H}_0 = \sum_{\alpha,i} E_i^{\alpha} c^{\dagger}_{\alpha,i} \hat{c}_{\alpha,i}
\end{equation}
where we account for the internal state dependent energies $E_i^{\alpha}$ in
motional state $i$.

The dipolar interactions in turn reduce to
\begin{equation}
     \hat{H}_{dd} =1/2 \sum_{ijkl} V_{ij}^{kl}
    \Bigg[ \sum_{\alpha,\alpha^{\prime}} \mu_{\alpha} \mu_{\alpha^{\prime}} \hat{c}_{\alpha k}^{\dagger} c^{\dagger}_{\alpha^{\prime} l} \hat{c}_{\alpha^{\prime} i} \hat{c}_{\alpha j} +  \mu_{\downarrow\uparrow} \mu_{\uparrow\downarrow}\left( \hat{c}_{\downarrow k}^{\dagger} \hat{c}_{\uparrow  l}^{\dagger} \hat{c}_{\downarrow i} \hat{c}_{\uparrow  j}  +h.c.\right)\Bigg]
\end{equation}
with the interaction matrix elements $V_{ij}^{kl}$ of the dipolar interactions in the chosen
single particle basis
\begin{equation}
V_{ij}^{kl} = \int d^2\vect{r} d^2\vect{r^{\prime}} \bar{\phi}_i(\vect{r}) \bar{\phi}_{j}(\vect{r}^{\prime}) V_{dd}(\vect{r}-\vect{r}^{\prime}) \phi_{k}(\vect{r}^{\prime}) \phi_l(\vect{r}) \, .
\end{equation}
Note that for simplicity we ignore the dependence of the wavefunctions on the
internal state $\alpha$ here, which will result in a slight renormalisation of
the matrix elements.

Finally, we will assume that the interactions are sufficiently weak, such that
they do not mix distinct single-particle states. Thus, we only retain terms in
the interactions that conserve the total single-particle energy, i.e the Hartree
and Fock terms. This leads to pairing the states as $l=i$, $j=k$, and $l=j$, $i=k$ for the
Hartree and Fock terms respectively.

Within these approximations we obtain
\begin{equation}
  \begin{split}
\hat{H}_{dd}&= 1/2 \sum_{ij}\Big[  (\mu_\downarrow^2\,  \hat{n}_{\downarrow i} \hat{n}_{\downarrow j}  +\mu_\uparrow ^2\, \hat{n}_{\uparrow i} \hat{n}_{\uparrow j} +2 \mu_\downarrow \mu_\uparrow   \,\hat{n}_{\downarrow i} \hat{n}_{\uparrow j}) V_{ij}^{ji} - \left( \mu_\downarrow^2\,  \hat{n}_{\downarrow i} \hat{n}_{\downarrow j}  +\mu_\uparrow ^2\, \hat{n}_{\uparrow i} \hat{n}_{\uparrow j}+\mu_\downarrow\mu_\uparrow  ( \hat{s}^{-}_{i} \hat{s}^{+}_{j} + \hat{s}^{+}_{i}\hat{s}^{-}_{j}) \right) V_{ij}^{ij} \\%
& \qquad \qquad +  2\mu_{\downarrow\uparrow}^2 \left(  \hat{s}^{-}_{i}\hat{s}^{+}_{j} V_{ij}^{ji} - \hat{n}_{\downarrow i}\hat{n}_{\uparrow j}V_{ij}^{ij} \right) \Big]
\end{split}
\end{equation}
Rewriting this in terms of the spin operators we obtain a long-range interacting XXZ type model in mode space given by
\begin{equation}
  \hat{H} = 1/2 \left(\sum_{i,j} \hat{s}^z_{i} \hat{s}^z_{j} J^z_{ij}  + 1/2\sum_{i,j} \left( \hat{s}^{-}_{i} \hat{s}^{+}_{j} + \hat{s}^{+}_{i}\hat{s}^{-}_{j} \right)  J^{\perp}_{ij}\right)+ \sum_{i} \hat{s}^z_{i} \left( \Delta E_i +h^z_i\right)
  \label{eq:spin_model_generic}
 \end{equation}
 where we dropped a constant energy shift and and $\Delta E_i = E_{\uparrow ,i}-E_{\downarrow ,i}$.
 The coupling constants are given in terms of the interaction matrix elements of
 the dipolar interactions in the single particle basis as
\begin{align}
  J^z_{ij}   &=  \left[ -\big( V_{ij}^{ij}-V_{ij}^{ji}\big) \big( \mu_\downarrow ^2+\mu_\uparrow ^2\big) -2 \mu_\downarrow  \mu_\uparrow  V_{ij}^{ji} +2 \mu_{\downarrow\uparrow}\mu_{\uparrow\downarrow} V_{ij}^{ij} \right] \label{app:couplings1} \\
  J^{\perp}_{ij}&=2  \left[ -\mu_\downarrow  \mu_\uparrow  V_{ij}^{ij}+\mu_{\downarrow\uparrow}\mu_{\uparrow\downarrow}  V_{ij}^{ji}\right] \label{app:couplings2} \\
   h^z_i   &=1/2\sum_{j}\left[ -\big( V_{ij}^{ij} -V_{ij}^{ji}\big) \big(\mu_\uparrow ^2-\mu_\downarrow ^2 \big) \right] \label{app:couplings3} 
\end{align}

\subsubsection{Harmonic oscillator space model}
In the harmonic oscillator basis the spin model is explicitly given as
\begin{equation}
  \hat{H} =1/2 \left( \sum_{\vect{n_1},\vect{n_2}} \hat{s}^z_{\vect{n_1}} \hat{s}^z_{\vect{n_2}} J^z_{\vect{n_1}\vect{n_2}}   + 1/2\sum_{\vect{n_1},\vect{n_2}} \left( \hat{s}^{-}_{\vect{n_1}} \hat{s}^{+}_{\vect{n_2}} + \hat{s}^{+}_{\vect{n_1}}\hat{s}^{-}_{\vect{n_2}} \right) J^{\perp}_{\vect{n_1},\vect{n_2}} \right)+ \sum_{\vect{n_1}} \hat{s}^z_{\vect{n_1}} (h^z_{\vect{n_1}} + \Delta E_{\vect{n}_1})
\end{equation}
with couplings
\begin{align}
  J^z_{\vect{n_1}\vect{n_2}}   &=  \Big[ -\big( V_{\vect{n_1}\vect{n_2}}^{\vect{n_1}\vect{n_2}}-V_{\vect{n_1}\vect{n_2}}^{\vect{n_2}\vect{n_1}}\big) \big( \mu_\downarrow ^2+\mu_{\uparrow}^2\big) -2 \mu_\downarrow  \mu_{\uparrow} V_{\vect{n_1}\vect{n_2}}^{\vect{n_2}\vect{n_1}} +2 \mu_{\downarrow\uparrow}\mu_{\uparrow\downarrow} V_{\vect{n_1}\vect{n_2}}^{\vect{n_1}\vect{n_2}} \Big] \label{app:ho_couplings1}\\
  J^{\perp}_{\vect{n_1}\vect{n_2}}&= 2 \left[ -\mu_\downarrow  \mu_{\uparrow} V_{\vect{n_1}\vect{n_2}}^{\vect{n_1}\vect{n_2}}+\mu_{\downarrow\uparrow}\mu_{\uparrow\downarrow}  V_{\vect{n_1}\vect{n_2}}^{\vect{n_2}\vect{n_1}}\right]\label{app:ho_couplings2}\\
   h^z_{\vect{n_1}}   &=1/2\sum_{\vect{n_2}}\left[ -\big( V_{\vect{n_1}\vect{n_2}}^{\vect{n_1}\vect{n_2}} -V_{\vect{n_1}\vect{n_2}}^{\vect{n_2}\vect{n_1}}\big) \big(\mu_{\uparrow}^2-\mu_\downarrow ^2 \big) \right] \label{app:ho_couplings3}
\end{align}

This basis has the distinct advantage of working in real space which makes it easy to include potential differences between the internal
levels. Unfortunately, the matrix elements of the interactions have to be evaluated numerically, making the interpretation of the couplings more difficult and limiting accessible system sizes, and thus temperature ranges and particle numbers.
However, we will derive below approximate expressions in the semi-classical limit  which allow us to understand the asymptotic scaling of the interactions.

\subsubsection{Interactions in mode space}
\begin{figure}
  \begin{minipage}[t]{0.99\columnwidth}
    \includegraphics[width=.99\columnwidth]{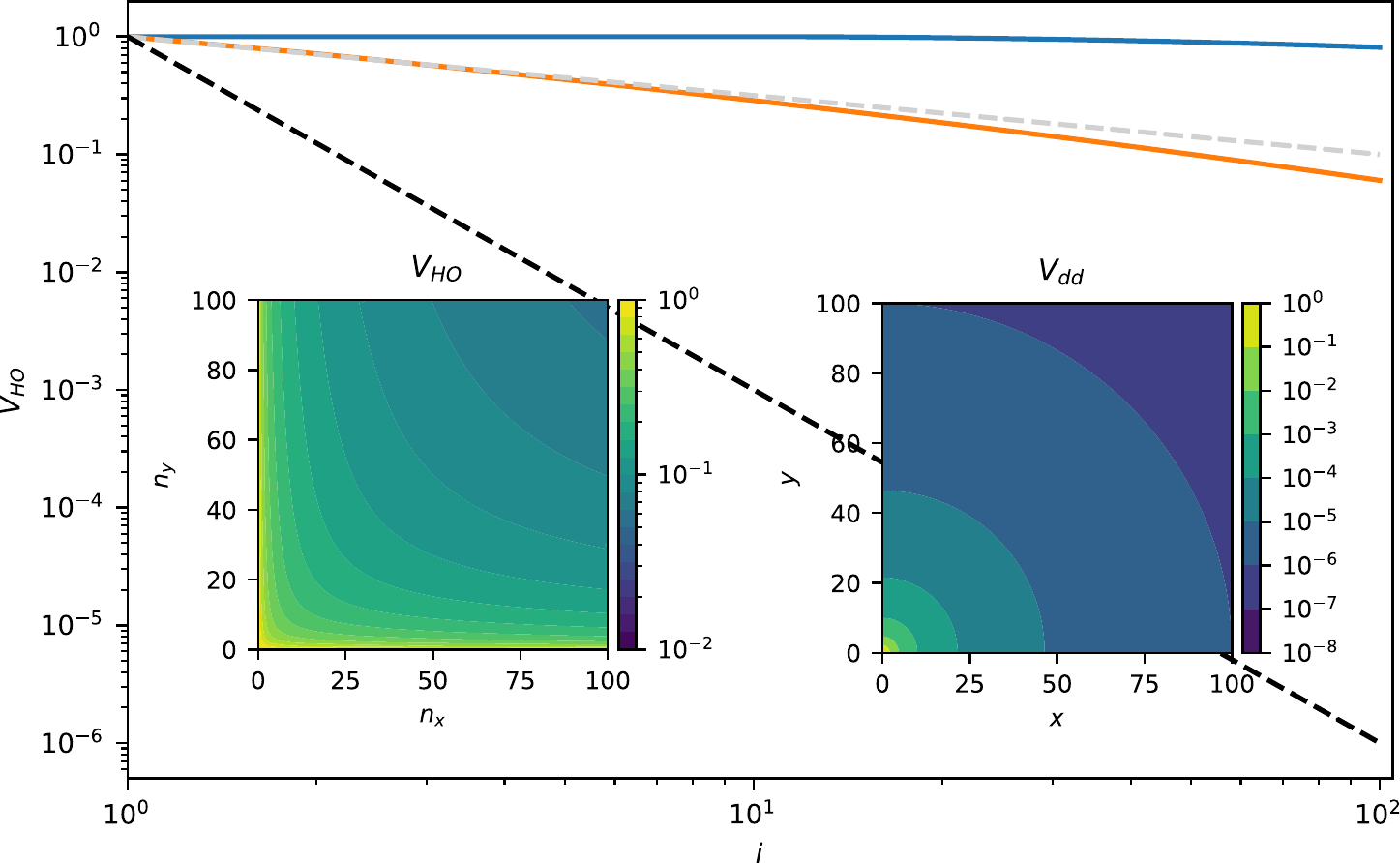}
  \end{minipage}
  \caption{Scaling of the interaction terms. Top: cuts of
    $V_{HO}(i,j)=V_{(0,0),(i,j)}^{(i,j),(0,0)}-V_{(0,0),(i,j)}^{(0,0),(i,j)}$
    along an edge $j=0$ (blue), and the diagonal $i=j$ (orange), dashed lines powerlaws in
    distance, $i^{-1/2}$ (grey), and $i^{-3}$ (black).
    Left inset: Harmonic oscillator mode space, full two-dimensional dependence of $V_{HO}(n_X,n_Y)$ %
    Right inset: Real space basis, $V_{dd}(X,Y)\sim (X^2+Y^2)^{-3/2}$
    \label{fig:HO_mode_space}}
\end{figure}
We next study form of the interactions of the spin-models in mode-space, and
contrast their behaviour to the real-space $1/R^3$ decay.
We characterise the interactions and their range by considering the behaviour of $V=V_{ij}^{ij}-V_{ij}^{ji}$ where the mode indices $i,j$  denote harmonic-oscillator modes $i=(n_X,n_Y)$.
The interactions in harmonic
oscillator mode space are shown in Fig.~\ref{fig:HO_mode_space} where we consider the
interactions of mode $(n_X,n_Y)$ with the mode $(0,0)$. We observe a weak
$i^{-1/2}$ (grey-dashed line) decay along the diagonal ($n_X=n_Y=i$), and almost constant
interactions along the edges ($n_X=0,n_Y=i$). Thus, the interactions are
considerably more longed-ranged in this basis than in real space (black dashed
lined and right inset) which is shown for comparison. In particular, note that
for the shown range, the interactions in mode space are 4 orders of magnitude
larger due to the more favourable scaling with distance.

\section{Computation of matrix elements in oscillator basis}
As there is no analytic expression for the matrix elements of the interaction in
the oscillator basis, we will in the following simplify these as much as
possible and reduce them to a semi-closed expression as sums over known
coefficients.

The interaction matrix element itself is given by
\begin{align}
   V_{\vect{n_1}\vect{n_2}}^{\vect{n_3}\vect{n_4}} &=  \int d^2 r d^2 r^{\prime} \bar{\phi}_{\vect{n_1}}(X^{\prime},Y^{\prime}) \bar{\phi}_{\vect{n_2}}(X,Y)   V^{2D}_{dd}(r-r^{\prime}) \phi_{\vect{n_3}}(X,Y)\phi_{\vect{n_4}}(X^{\prime},Y^{\prime}))\\
    &=\frac{1}{(2 \pi)^2}  \int d^2 q \mathcal{F}\left[\phi_{\vect{n_1} }\phi_{\vect{n_4}} \right](q) \mathcal{F}\left[\phi_{\vect{n_2}} \phi_{\vect{n_3}} \right](-q)  V(q)\\
\end{align}
where
\begin{equation}
    \mathcal{F}\left[\phi_{\vect{n_1} }\phi_{\vect{n_2}} \right](q) = \int d^2 r e^{i q r} \phi_{n_{1,X}}(X)\phi_{n_{1,Y}}(Y) \phi_{n_{2,X}}(X) \phi_{n_{2,Y}}(Y)
\end{equation}
is the Fourier-Transform of the product of HO wavefunctions. This integral of course separates over $(X,Y)$, $(q_X,q_Y)$ respectively, since the wavefunction separates. However, the interaction integral itself does not separate since the interaction depends on $|q|$ and does not separate in $(q_X,q_Y)$.

Explicitly the one-dimensional integrals are given by
\begin{equation}
    \int d X e^{i q X} \phi_{n_{1}}(X)\phi_{n_{2}}(X) = \sqrt{\frac{2^{n_2} n_2!}{2^{n_1}n_1!}} e^{i \pi(n_1-n_2)/2} e^{-a_{\perp}^2 q^2/4} \left(-\frac{a_{\perp} q}{2}\right)^{n_2-n_1} L_{n_1}^{n_2-n_1}(a_{\perp}^2q^2/2)
\end{equation}
where $L_n^m$ is an associated Laguerre polynomial, and $a_{\perp}=\sqrt{\hbar/(m \omega_{\perp})}$ is the harmonic oscillator length in the transverse/plane direction.

\subsubsection{Quasi-closed expression}
The integral for the interaction matrix elements thus contains terms of the form
\begin{equation}
L_{r_1}^{s_1}(q \cos(\phi)) L_{r_2}^{s_2}(q \cos(\phi)) L_{r_3}^{s_3}(q \sin(\phi)) L_{r_3}^{s_3}(q \sin(\phi))
\end{equation}
Multiplying the polynomials out we get integrals of the type
\begin{equation}
  I(n_1,n_2)= \int d \phi \sin(\phi)^{n_1} \cos(\phi)^{n_2} = \big( 1+(-1)^{n_2} \big) \big( 1+(-1)^{n_1+n_2} \big) \frac{\Gamma\left(\frac{1+n_1}{2}\right)\, \Gamma\left(\frac{1+n_2}{2}\right)}{2 \Gamma\left(\frac{2+n_1+n_2}{2}\right)}
\end{equation}
where $\Gamma(x)$ is the Gamma function.

Each such term also has a factor $q^{n_1+n_2}$, leading to the remaining integral
\begin{equation}
    \begin{split}
   J(n_1,n_2)=  \int dq\, q q^{n_1+n_2} V(a_z/a_{\perp} q) e^{-q^2/2} &= \frac{1}{3} 2^{-\frac{n}{2}-
    \frac{3}{2}} \Bigg(\frac{2^{n+3} \Gamma \left(\frac{n}{2}+1\right)}{\sqrt{\pi }}\\
    & \qquad -3 c_1^{-n-2} \Gamma (n+3) \, _2\tilde{F}_1\left(\frac{n+3}{2},\frac{n+4}{2};\frac{n+5}{2};1-\frac{1}{\hat{c}_1^2}\right)\Bigg)
    \end{split}
\end{equation}
where $n=n_1+n_2$ and $c_1=a_Z/a_{\perp}$, and $_2\tilde{F}_1$ is the regularized hypergeometric function

Thus, the full integral is given by sums over these terms multiplied by the coefficients of the Laguerre polynomials at the corresponding order.

\subsubsection{Hartree Term}
To illustrate the closed form expression motivated above, we here provide explicit expressions for the Hartree term $V_{\vect{n_1}\vect{n_2}}^{\vect{n_2}\vect{n_1}}$.

For brevity we explicitly define the m-th polynomial coefficient of the associated Laguerre Polynomial $L_n^k(q)$, e.g. the prefactor of the $q^m$ term, as
\begin{equation}
    f^k_n(m) = \frac{(-1)^m (k+n)!}{(m! (k+m)! (n-m)!)}
\end{equation}
and we also define $f_n(m) = f_n^{k=0}(m)$ which is the corresponding coefficient for the usual Laguerre polynomials.

In total, we get
\begin{equation}
\begin{split}
 V_{(n_{1X},n_{1Y}),(n_{2X},n_{2Y})}^{(n_{2X},n_{2Y}),(n_{1X},n_{1Y})}&= \sum_{l_X=0}^{n_{1X}+n_{2X}}\sum_{l_Y=0}^{n_{1Y}+n_{2Y}} %
 I(2 l_X,2 l_Y) \, J(2 l_X+2 l_Y)
 \sum_{\substack{0 \le m_{1X}\le n_{1X}\\0\le m_{2X}\le n_{2X}\\m_{1X}+m_{2X}=l_X}} \frac{f_{n_{1X}}(m_{1X}) f_{n_{2X}}(m_{2X})}{2^{m_{1X}} 2^{m_{2X}}}\\
& \quad \times  \sum_{\substack{0 \le m_{1Y} \le n_{1Y}\\0 \le m_{2Y} \le n_{2Y}\\m_{1Y}+m_{2Y}=l_Y}} \frac{f_{n_{1Y}}(m_{1Y}) f_{n_{2Y}}(m_{2Y})}{2^{m_{1Y}} 2^{m_{2Y}}}
\end{split}
\end{equation}
where $I(n_1,n_2)$ and $J(n)$ are the integrals defined above.

Note that to evaluate these sums still requires extremely high numerical precision since
the individual terms have alternating signs and a magnitude growing extremely fast with the
maximal oscillator number, while the total sum is of order 1 and decreasing with the oscillator number.

\section{Semi-classical calculation}
In addition to the exact calculation presented above, we also consider a semi-classical calculation to understand the scaling behaviour of the interaction matrix elements.

Starting from the direct (Hartree) integral
\begin{align}
    V_{d} &= \int d^3 r \int d^3 r^{\prime} \, \left| \phi_i\right|^2(r)  V(r-r^{\prime}) \left| \phi_j\right|^2(r^{\prime}) 
\end{align}
and making use of the Wigner transform 
\begin{equation}
    W(r,k) = \int d^3 r^{\prime}\,  \bar{\hat{\psi}}(r+r^{\prime}/2) \hat{\psi}(r-r^{\prime}/2) e^{i k r}
\end{equation}
normalised as $\int d^3r \int \frac{d^3k}{(2\pi)^3}\, W(r,k) =1$
we can rewrite this as
\begin{align}
    V_{d} &= \int \frac{d^3k}{(2 \pi)^3} \int \frac{d^3k^{\prime}}{(2 \pi)^3}\int \frac{d^3k^{\prime\prime}}{(2 \pi)^3}    \tilde{W}_i(-k^{\prime \prime},k) \tilde{W}_j(k^{\prime \prime},k^{\prime})  \, \tilde{V}(k^{\prime \prime})
\end{align}
where $\tilde{W}(k_1,k_2)  = \int d^3r \, e^{-i r k_1} W(r,k_2)$ and $\tilde{V}(k) =\int d^3r \, e^{-i r k} V(r)$ are the Fourier-transform of the Wigner function and of the three-dimensional dipolar interactions respectively.

For the one-dimensional harmonic oscillator the classical limit of the Wigner function is given by \cite{Truman1999}
\begin{equation}
    W^{(1D)}_E(X,k_X) = A \delta \left(E -  \frac{m \omega^2 X^2}{2} -  \frac{\hbar^2k_X^2}{2m} \right)
\end{equation}
with the normalisation factor $A=4 \hbar \omega$. 

Thus, for a 3D separable potential we have
\begin{equation}
    W^{(3D)}_{E_x,E_y,E_z}(X,Y,Z,k_X,k_Y,k_Z) = \prod_i A_i \delta \left(E_i -  \frac{m \omega_i^2 X_i^2}{2} -  \frac{\hbar^2k_i^2}{2m} \right)
\end{equation}

We approximate this further as
\begin{equation}
    W^{(3D)}_E(r,k) = A \delta \left(E - \sum_i \frac{m \omega_i^2}{2} r_i^2 - \sum_i \frac{\hbar^2 k_i^2}{2m} \right)
\end{equation}
with the normalisation constant $A= \frac{2 (\hbar \omega)^3}{E^2}$ with $\omega = (\omega_X\omega_Y\omega_Z)^{1/3}$.

Now using $V(r) = \frac{1}{ 4\pi r^3} (1-3\cos(\theta)^2) $, $\tilde{V} =\cos(\theta_k)^2-1/3 $, we can solve the integrals analytically to obtain
\begin{align}
   V_d &=N_d \sqrt{2 E_2/(\hbar\omega)} g(\sqrt{E_1/E_2}) G(\omega_Z/\omega_X,\omega_Z/\omega_Y) 
\end{align}  
assuming that $E_2>E_1$  with $N_{d} =\frac{1}{(2\pi)^3} \frac{16}{E_1/(\hbar\omega) E_2/(\hbar\omega)} \frac{1}{a_{ho}^3}$ 
 and
\begin{equation}
    g(b) = \frac{16 \left(b^4-b^2+1\right) E\left(b^2\right)-8 \left(b^4-3 b^2+2\right) K\left(b^2\right)}{15 \pi  b^2}
\end{equation}
where $K$ and $E$ are complete elliptic functions of the first and second kind, and the anisotropy function $G$ is defined as
\begin{equation}
  G(a,b)=\frac{4\pi}{3} \left(\frac{3 ab (E(\phi,k)-F(\phi,k))}{\sqrt{1-a^2} \left(1-b^2\right)}+1\right)
\end{equation}
with $\phi= \arcsin(\sqrt{1-a^2})$, $k=(1-b^2)/(1-a^2)$ and $E(\phi,k)$ ($F(\phi,k)$) are the incomplete elliptic integrals of the first (second) kind.

We note that based on this result we have $V_d \sim \frac{\sqrt{\max\{E_1,E_2\}}}{E_1 E_2}$
implying $V\sim 1/E^{3/2}$ for $E_1=E_2=E$ and $V\sim 1/\sqrt{E_2}$ for $E_1=const$ and $E_2>E_1$.
To connect with the discussion of mode space lattices, note that since $E= i \hbar \omega$ for mode $i$, this translates into a scaling $V \sim 1/i^{3/2}$ and $V\sim 1/\sqrt{i}$ respectively.

\section{Spin self-rephasing within kinetic model\label{app:kinetic}}
To understand the relevance of collisions on the spin dynamics, in particular,
collision-induced decoherence, we consider a simple kinetic model adopted from Ref.~\cite{Deutsch_2010}. 
In this approach the dynamics of the spin density $\vect{S}(E)$ as a function of the energy $E$ is described by

\begin{equation}
\partial_t \vect{S}(E,t) +\gamma \left[ \vect{S}(E,t) - \bar{\vect{S}}(t) \right] \approx \left[ \Delta(E) \vect{e}_z +\int_0^{\infty} dE\, V(E,E^{\prime}) \vect{S}(E^{\prime},t))\right] \times \vect{S}(E,t)
\end{equation}
where $\gamma$ describes the relaxation of the spin-density due to collisions to its average
$\bar{\vect{S}}(t) = \int dE \rho(E) n_F(E) \vect{S}(E,t)$ with $\rho(E)$ the
density of states, and $n_F(E)$ the Fermi-Dirac distribution, $\Delta(E)=
\frac{\omega_1-\omega_0}{(\omega_0+\omega_1)/2} E$ accounts for the different trapping
potentials of the internal states, and $V(E,E^{\prime})$ is the matrix-valued
interaction kernel given in terms of
Eqs~\ref{app:couplings1}--\ref{app:couplings3} or
Eqs~\ref{app:ho_couplings1}--\ref{app:ho_couplings3} respectively.
Without interactions the single-particle term $\Delta(E)$ will lead to quick dephasing and decay of the collective spin over a time-scale set by the variance of $\Delta(E)$.
We defer the computation of the relaxation rate $\gamma$ to the next
section.

\begin{figure}
  \begin{minipage}[t]{0.99\columnwidth}
    \includegraphics[width=.99\columnwidth]{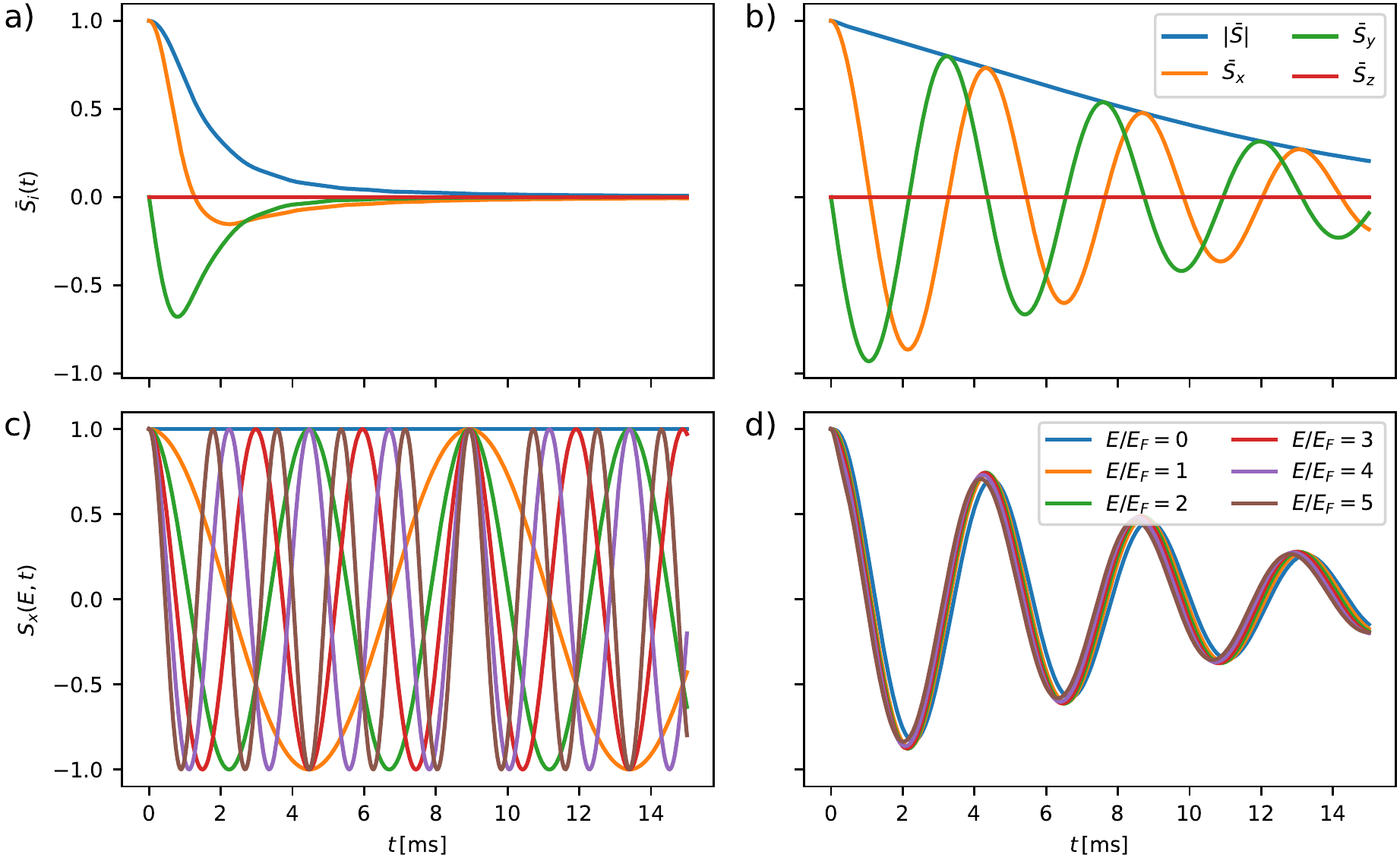}
  \end{minipage}
    \caption{Spin dynamics within the kinetic theory model for a system of $N=1000$
      particles at $T/T_F = 1.0$. Left column non-interacting with dephasing
      $\Delta \omega/\omega=0.05$, right column
      including dipolar interactions, dephasing and relaxation with
      $\gamma=5000\, 1/s$. Panels a), b) show the dynamics of the
    $x,y,z$ components of the average
    spin-density $\bar{\vect{S}}$, and its norm $\left| \bar{\vect{S}} \right|$.
    Panels c), d) show the dynamics of $S_x(E_i,t)$ for a range of different
    energies again comparing the non-interacting (left) to the interacting (right) system. 
    \label{fig:kinetic_theory_dyn}}
\end{figure}

We show the resulting dynamics in Fig.~\ref{fig:kinetic_theory_dyn} comparing
the non-interacting theory (left column) to the interacting dynamics (right
column). Note that in the non-interacting case for the chosen parameters the collective spin decays on a
time-scale of $t \approx 5$ ms due to single-particle dephasing $(\Delta(E))$. In contrast,
this decay is significantly suppressed by the dipolar interactions. In fact, we
observe that the individual energy levels stay quite coherent throughout the dynamics, and the
observed decay of the collective spin is due to the interaction induced loss of coherence ($\gamma$) over a
significantly longer timescale instead.

\section{Computation of spin-relaxation time within kinetic theory}
To include effects of collisions not included in the spin-model we employ a kinetic approach to obtain
a relaxation rate of the spin-degrees of freedom.

Following Refs.~\cite{Bruun_2012,Schafer_2012,Enss_2012,Baur_2013} we describe the semi-classical dynamics of the
two-dimensional Fermi gas via the Boltzmann equation
\begin{equation}
\left( \partial_t + \frac{p}{m} \cdot \nabla_r -\nabla_r U(r) \cdot \nabla_p \right) f_{\alpha} = - I[f_{\alpha}]
\end{equation}
for the distribution function $f_{\alpha}(r,p)$ of internal state $\alpha$.

We now linearise the Boltzmann equation around the (non-interacting) equilibrium
solutions
\begin{equation}
f^{(0)}(r,p) = \frac{1}{e^{\beta \left(\frac{p^2}{2m} +U(r) - \mu\right)}+1}
\end{equation}
as $f(r,p) = f^{0}(r,p) +f^{(0)}(r,p)\left[ 1 - f^{(0)}(r,p) \right]
\phi_{\sigma}(r,p)$.
Since we are interested in the relaxation of the spin-degrees of freedom due to
collisions, we choose $\phi_{\sigma} = \pm p_X$ with opposite signs for the
different internal states.

The linearised Boltzmann equation reads as
\begin{equation}
f^{(0)}(1-f^{(0)}) \left( \partial_t + \frac{p}{m} \cdot \nabla_r -\nabla_r U(r) \cdot \nabla_p \right) \phi_{\sigma} = - I[\phi_{\sigma}]
\end{equation}
and the collision integral is given by
\begin{equation}
I[\phi_{\sigma}] = \int \frac{d^2p}{(2 \pi )^2} \frac{m_r}{2} \int_0^{2 \pi} d\theta^{\prime} \left| \mathcal{T} \right|^2 f^{0}_{1,\sigma} f^{0}_{2,\bar{\sigma}} (1-f^{(0)}_{3,\sigma})(1-f_{4,\bar{\sigma}}^{(0)}) \left( \phi_{1,\sigma} + \phi_{2,\bar{\sigma}} -\phi_{3,\sigma} -\phi_{4,\bar{\sigma}}\right)
\end{equation}
where $1,2$ denote the particles before the collision $3,4$ the particles after
the collision $f_i=f(r,p_i)$, etc. The T-matrix $\mathcal{T}$ describes the
scattering process of particles $1$ and $2$ with momenta $p_1$, $p_2$, relative
momentum $p_r=(p_2-p_1)/2$ and center
of mass momentum $P=p_1+p_2$ scattering to states $3,4$ with momenta $p_3$, $p_4$
under conservation of total momentum, where $\theta^{\prime}$ describes the
angle between $P$ and the final relative momentum $p_{r}^{\prime}=(p_4-p_3)/2$.
We use the low-energy form of the T-matrix for $s$-wave scattering in 2D given
by
\begin{equation}
\mathcal{T}(\epsilon_{p_r}) = \frac{2 \pi}{m_r} \frac{1}{\ln(\epsilon_b/\epsilon_{p_r}) + i \pi} 
\end{equation}
depending on the kinetic energy of the relative motion of colliding particles
$\epsilon_{p_r} = \frac{p_r^2}{2m_r}$ and the bound state energy $\epsilon_b =
\frac{\hbar^2}{m a_{2D}}$. The two-dimensional s-wave scattering length in the quasi-2D limit
is given as $a_{2D} = a_z \sqrt{\pi/B} e^{-\sqrt{\pi/2}a_z/a_{3D}}$ in terms of
the 3D scattering length $a_{3D}$ and the harmonic oscillator length $a_z =
\sqrt{\hbar/(m \omega_z)}$ and $B\approx 0.905$.

Finally, we use the relaxation time approximation writing $I[\phi] \approx
\phi/\tau$, and average the linearised Boltzmann equation to get

\begin{equation}
  \frac{1}{\tau} = \frac{\int d^2r d^2p \, \phi_{\sigma} I[\phi_{\sigma}]}{\int d^2r d^2p \, \phi_{\sigma}^2}
\end{equation}

Using the symmetries of the collision integral we can slightly simplify the
enumerator to
\begin{equation}
  m_r \int \frac{d^2P}{(2 \pi)^2} \frac{d^2p_r}{(2\pi)^2} d^2r d\theta d\theta^{\prime} \left| \mathcal{T} \right|^2 p_r^2 (1-\cos(\theta-\theta^{\prime})) f^{0}_{1} f^{0}_{2} (1-f^{(0)}_{3})(1-f_{4}^{(0)})
\end{equation}
The denominator can integrated analytically as
\begin{equation}
\int \frac{d^2p}{(2\pi )^2} d^2r \, p_X^2 f^{(0)}(1-f^{(0)}) = -\frac{m}{\beta^3 \omega^2} \mathrm{Li}_2(-z)
\end{equation}
where $\mathrm{Li}_2$ is a polylogarithm.

\begin{figure}
  \begin{minipage}[t]{0.49\columnwidth}
    \includegraphics[width=.99\columnwidth]{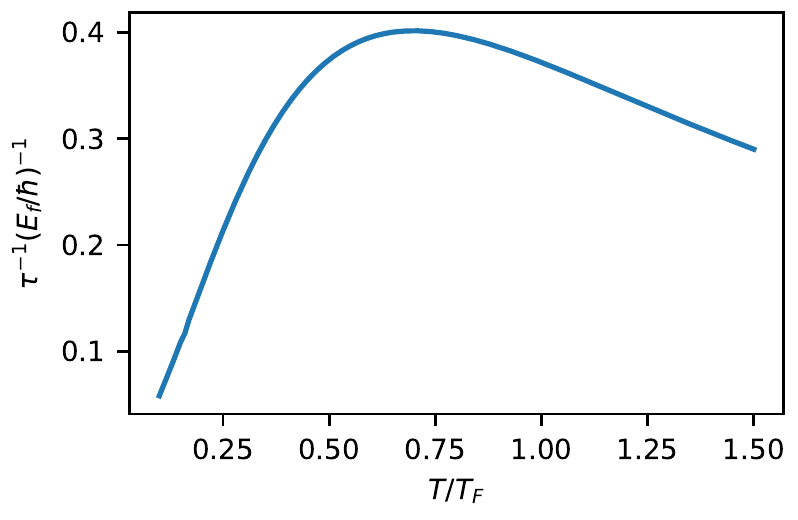}
  \end{minipage}
  \caption{Spin relaxation time $\tau$ in units of the Fermi-Energy $E_F$ versus
    temperature in units of $T_F$ in the $s$-wave approximation, for realistic
    experimental parameters, $N=1000$, $\omega_Z = 20$ kHz.
    \label{fig:Vq}}
\end{figure}

Using these expressions, we can numerically obtain the spin-relaxation time. To
make connection with the dipolar quasi-2D gas we approximate the collisions due to the
dipolar interactions between different internal states as $s$-wave scattering
with scattering length $a_{3D} = \sqrt{\sigma_{dd}/(4\pi)}$ from the universal scattering cross
section $\sigma_{dd} =  32 \pi/15 a_{dd}^2$ , where $a_{dd} = \frac{(\mu_{\downarrow\uparrow}
  d_0)^2}{8 \pi \epsilon_0} \frac{M}{\hbar^2}$ is the dipole-length \cite{Bohn_2009}.
The resulting spin relaxation time as a function of temperature is shown in Fig.~\ref{fig:Vq} for $N=1000$ particles confined by a $\omega_z=20$ kHz harmonic potential along $z$. 

\section{S-wave losses}
Finally, we consider the inclusion of lossy s-wave collision in our description. While collisions of particles in the same state, e.g. particles in $\alpha=0$, can only occur in the p-wave (and higher angular momentum) channel, which is suppressed at low temperatures by the centrifugal barrier, a molecule in $\alpha = 0$ can collide with a molecule in $\alpha = 1,\tilde{1}$ via s-wave collisions without  such suppression. 
Note  that s-wave collisions are still forbidden as long as the molecules remain in the same collective spin-state. Therefore, these s-wave losses are only relevant if the system dephases. As we have shown above,  dipolar interactions can generate   strong spin-locking thanks to  the many-body gap which suppresses dephasing and therefore s-wave losses as well.

To model these s-wave losses, we consider the evolution of the full density matrix $\rho$ evolving according to
\begin{equation}
    \dot{\rho} =- \frac{i}{\hbar}\left[\rho, H \right] +1/2\sum_{ijkl} \Gamma_{ij}^{kl} \left( A_{ij} \rho A_{kl} -1/2 \left\{A^{\dagger}_{ij} A_{kl},\rho \right\} \right)
\end{equation}
where $H$ is the dipolar Hamiltonian of the elastic interaction already discussed and $A_{ij}$ are jump operators modelling the s-wave collisional loss as $A_{ij} = c_{e,i} c_{g,j}$, with $g,e$ denoting the states $\alpha=0$ or $\alpha=1,\tilde{1}$ respectively.
The coefficients $\Gamma_{ij}^{kl}$ are given by $s$-wave integrals over the harmonic oscillator wave-functions as
\begin{equation}
    \Gamma_{ij}^{kl} = \frac{4 \pi \hbar a_{sc}}{M} \prod_{r=X,Y,Z} \int dr  \int dr^{\prime} \, \bar{\phi}_{n_{i,r}}(r) \bar{\phi}_{n_{j,r}}(r) \delta(r-r^{\prime}) \phi_{n_{k,r}}(r^{\prime}) \phi_{n_{l,r}}(r^{\prime})
\end{equation}
with $a_{sc}$ the $s$-wave scattering length. Following Ref.~\cite{PhysRevLett.105.073202} we assume the s-wave losses to be determined universally directly by the s-wave scattering length of KRb, $a_{sc}=118 a_0$ in term of the Bohr radius. We note that these integrals can be analytically expressed in closed form.

Starting from the master equation we can now derive equations of motion for the observables in the spin model, similar to Ref. \cite{Rey2014}, and we again use the no-mode changing collision approximation and a mean-field decoupling.
We consider the evolution of the components of the density matrix given by, $\rho_{ee}^{mm} =\langle\hat{c}^{\dagger}_{em} \hat{c}_{em}\rangle$, $\rho_{gg}^{mm} =\langle\hat{c}^{\dagger}_{gm} \hat{c}_{gm}\rangle$, $\rho_{eg}^{mm} =\langle\hat{c}^{\dagger}_{em} \hat{c}_{gm}\rangle$ and $\rho_{ge}^{mm} =\langle\hat{c}^{\dagger}_{gm} \hat{c}_{em}\rangle$.

For completeness we first reproduce the equations of motion stemming from the Hamiltonian part
\begin{align}
   \Big( \frac{d \rho^{mm}_{ee}}{dt}\Big) _H &=-i\sum_i (\rho_{ge}^{ii} \rho_{eg}^{mm}-\rho_{eg}^{ii} \rho_{ge}^{mm}) \left(\mu_{eg}\mu_{ge} V_{im}^{mi}-\mu_{g} \mu_{e} V_{im}^{im}\right)\\
  \Big( \frac{\rho^{mm}_{gg}}{dt} \Big)_H &=i \sum_i (\rho_{ge}^{ii} \rho_{eg}^{mm}-\rho_{eg}^{ii} \rho_{ge}^{mm}) \left(\mu_{eg}\mu_{ge} V_{im}^{mi}-\mu_{g} \mu_{e} V_{im}^{im}\right)\\
   \Big(\frac{d \rho^{mm}_{eg}}{dt}\Big)_H &=i \sum_i \mu_{g}^2 (V_{im}^{mi} - V_{im}^{im})\rho_{gg}^{ii} \rho_{eg}^{mm} -\mu_{e}^2 (V_{im}^{mi} -V_{im}^{im})\rho_{ee}^{ii} \rho_{eg}^{mm}\\
&\quad +\rho_{eg}^{ii} \rho_{ee}^{mm} \left(\mu_{eg}\mu_{ge} V_{im}^{mi}-\mu_{g} \mu_{e} V_{im}^{im}\right)+\rho_{eg}^{ii} \rho_{gg}^{mm} \left(\mu_{g} \mu_{e} V_{im}^{im}-\mu_{eg}\mu_{ge} V_{im}^{mi}\right)\\
&\quad +(-\mu_{g} \mu_{e} V_{im}^{mi} +\mu_{eg}\mu_{ge} V_{im}^{im}) \rho_{gg}^{ii} \rho_{eg}^{mm}+(\mu_{g} \mu_{e} V_{im}^{mi}-\mu_{eg}\mu_{ge} V_{im}^{im}) \rho_{ee}^{ii} \rho_{eg}^{mm}
\end{align}


The lossy s-wave collisions result in the following contribution to the equations of motion
\begin{align}
    \frac{d \rho^{mm}_{ee}}{dt} &= \Big( \frac{d \rho^{mm}_{ee}}{dt}\Big) _H +\frac{1}{2} \rho_{ge}^{mm}\sum _i \gamma_{im} \rho_{eg}^{ii}+ \frac{1}{2}\rho_{eg}^{mm} \sum_i \gamma_{mi} \rho_{ge}^{ii}  -\rho_{ee}^{mm}\sum _i \gamma_{im} \rho_{gg}^{ii}\\
   \frac{\rho^{mm}_{gg}}{dt} &=\Big( \frac{d \rho^{mm}_{gg}}{dt}\Big) _H+ \frac{1}{2} \rho_{ge}^{mm}\sum _i \gamma_{im} \rho_{eg}^{ii}+ \frac{1}{2}\rho_{eg}^{mm} \sum_i \gamma_{mi} \rho_{ge}^{ii}  -\rho_{gg}^{mm}\sum _i \gamma_{im} \rho_{ee}^{ii}\\
   \frac{d \rho^{mm}_{eg}}{dt} &=\Big( \frac{d \rho^{mm}_{eg}}{dt}\Big) _H+\frac{1}{2}(\rho_{ee}^{mm}+ \rho_{gg}^{mm}) \sum_i \gamma_{im} \rho_{eg}^{ii}  -\frac{1}{2} \rho_{eg}^{mm} \sum_i \gamma_{im}(\rho_{ee}^{ii}+\rho_{gg}^{ii})
\end{align}
where we defined $\gamma_{im} = \Gamma_{im}^{im} = \Gamma_{im}^{mi}$ for $s$-wave interactions.

Rewriting the equations of motion in terms of
$s^x_m = (\rho^{mm}_{eg}+\rho^{mm}_{ge})/2$, $ s^y_m = (\rho^{mm}_{eg}+\rho^{mm}_{ge})/(2 i)$, $ s^z_m = (\rho^{mm}_{ee}-\rho^{mm}_{gg})/2$ and $\rho_m = (\rho^{mm}_{ee}+\rho^{mm}_{gg})$ 
we obtain for the Hamiltonian part
\begin{align}
\Big(\frac{ds^x_m}{dt}\Big)_H &=\sum_i s^z_i s^y_m \left( (\mu_g^2+\mu_e^2)(V_{im}^{mi}-V_{im}^{mi} ) - 2\mu_e \mu_g V_{im}^{mi}+2\mu_{eg}\mu_{ge} V_{im}^{im} \right)+ s^y_i s^z_m \left(2\mu_g \mu_e V_{im}^{im}-2\mu_{eg}\mu_{ge} V_{im}^{mi}\right)\\
    &\quad -\frac{1}{2} s^y_m (\mu_g^2-\mu_e^2)(V_{im}^{mi}-V_{im}^{im}) \rho_i  \\
\Big(\frac{ds^y_m}{dt}\Big)_H &=\sum_i -s^z_i s^x_m \left( (\mu_g^2+\mu_e^2)(V_{im}^{mi}-V_{im}^{mi} ) - 2\mu_e \mu_g V_{im}^{mi}+2\mu_{eg}\mu_{ge} V_{im}^{im} \right)- s^x_i s^z_m \left(2\mu_g \mu_e V_{im}^{im}-2\mu_{eg}\mu_{ge} V_{im}^{mi}\right)\\
    &\quad +\frac{1}{2}  s^x_m (\mu_g^2-\mu_e^2)(V_{im}^{mi}-V_{im}^{im}) \rho_i \\
\Big (\frac{ds^z_m}{dt} \Big )_H&= \sum_i  \left(2\mu_{eg}\mu_{ge} V_{im}^{mi}-2\mu_g \mu_e V_{im}^{im}\right) (s^x_i s^y_m-s^y_i s^x_m)\\
\Big(\frac{d \rho_m}{dt}\Big)_H  &= 0
\end{align}
Note that the density per mode $\rho_m$ has explicitly no dynamics in the Hamiltonian evolution due to the no-mode-changing collisions approximation. Thus, we recover the equations of motion obtained from the spin-hamiltonian, Eq.~\ref{eq:spin_model_generic}, with the only modification that the $z$ field, previously defined in Eq.~\ref{app:couplings3}, is now given by
\begin{equation}
   h^z_i   =1/2\sum_{j}\left[ -\big( V_{ij}^{ij} -V_{ij}^{ji}\big) \big(\mu_e^2-\mu_g^2 \big) \right] \rho_j
\end{equation}
which explicitly depends on $\rho_j$.

The lossy s-wave collisions contribute in addition
\begin{align}
\frac{ds^x_m}{dt} &= \Big (\frac{ds^x_m}{dt} \Big )_H+1/2 \left(\rho_m \sum_{i} \gamma_{mi} s^x_i - s^x_m \sum_{i} \gamma_{mi} \rho_i \right) \\
\frac{ds^y_m}{dt} &= \Big (\frac{ds^y_m}{dt} \Big )_H+ 1/2 \left(\rho_m \sum_{i} \gamma_{mi} s^y_i - s^y_m \sum_{i} \gamma_{mi} \rho_i \right) \\
\frac{ds^z_m}{dt} &=\Big (\frac{ds^z_m}{dt} \Big )_H+ 1/2 \left(\rho_m \sum_{i} \gamma_{mi} s^z_i - s^z_m \sum_{i} \gamma_{mi} \rho_i \right) \\
\frac{d \rho_m}{dt}  &= \Big (\frac{d\rho_m}{dt} \Big )_H+ 1/2 \left(4  \left(s_m^x \sum_i \gamma_{mi} s_i^x+s_m^y \sum_i \gamma_{mi}s_i^y+s_m^z \sum_i \gamma_{mi} s_i^z\right)  -\rho_m \sum_i \gamma_{mi} \rho_i \right) 
\end{align}
where now the density per mode $\rho_m$ acquires dynamics.
However, we note that for a fully collective state the equation for $\rho_m$ vanishes identically, and the density is also conserved, as expected due to the s-wave nature of the collisions.

\subsection{Decay dynamics}
\begin{figure}
  \begin{minipage}[t]{0.99\columnwidth}
    \includegraphics[width=.99\columnwidth]{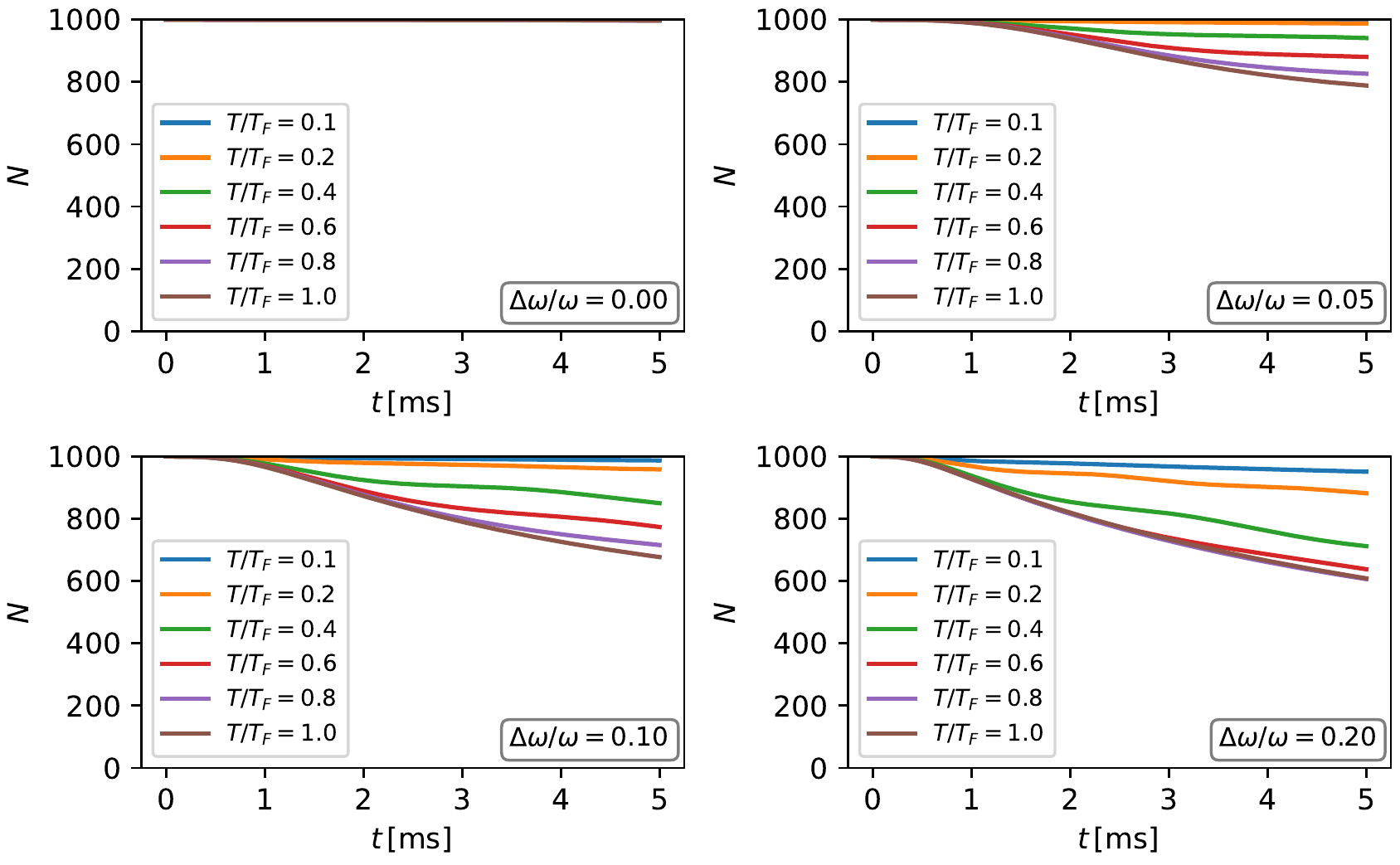}
  \end{minipage}
    \caption{Dynamics of the total number of molecules $N$ versus time $t$ in presence of s-wave losses at different temperatures $T/T_F =0.1, \cdots, 1.0$ and different single-particle dephasing strengths $\Delta_{\omega}$ as indicated in the panels. Starting from an initially coherent spin state of $N_0=1000$ molecules polarized along $x$.
    \label{fig:loss_dynamics}}
\end{figure}

Solving the generalised spin-model including the dipolar interactions and the s-wave losses, allows us to answer whether the many-body gap protecting from single-particle dephasing in the spin model also suppresses the s-wave losses for the time-scales of interest.

Based on this expecation we study the dynamics of the total particle number in the spin model as a function of the single-particle dephasing $\Delta_{\omega}$ and temperature $T$ in Fig.~\ref{fig:loss_dynamics}. Indeed for vanishing dephasing $\Delta_{\omega}=0$ (upper left panel) we observe no relevant particle loss. In contrast for the largest dephasing $\Delta_{\omega}=0.2$ there is significant particle loss, which is successively reduced at lower dephasing strength. We also note that the particle loss decreases with the temperature of the gas. Most importantly, the squeezing time is on the order of 1.5 ms at the lowest temperatures to about 2 ms at the largest temperatures here. Thus, the particle loss can be fully suppressed for the time-scales of interest if the temperature is low enough even for finite dephasing strength.

%